\documentclass[pre,showpacs,twocolumn,preprintnumbers,amsmath,amssymb,superscriptaddress]{revtex4}
\usepackage{graphicx}
\usepackage{dcolumn}
\usepackage{bm}
\usepackage{natbib}
\usepackage[nativepdf,hyperfigures=false,colorlinks=true,citecolor=blue,linkcolor=black,anchorcolor=black,urlcolor=black]{hyperref}
\usepackage[absolute]{textpos}

\newcommand{\ER}{Erd\H{o}s-R\'{e}nyi}
\newcommand {\bfR} {$\boldsymbol{\rho}^\mathrm{c}$}
\newcommand {\bfRm} {$\boldsymbol{\rho}^\mathrm{m}$}
\newcommand {\CCO} {$C_\mathrm{c}$} 
\newcommand {\LCO} {$L_\mathrm{c}$} 
\newcommand {\CMO} {$C_\mathrm{m}$} 
\newcommand {\LMO} {$L_\mathrm{m}$} 
\newcommand {\CCa} {$C^{(1)}_\mathrm{c}$} 
\newcommand {\LCa} {$L^{(1)}_\mathrm{c}$} 
\newcommand {\CMa} {$C^{(1)}_\mathrm{m}$} 
\newcommand {\LMa} {$L^{(1)}_\mathrm{m}$} 
\newcommand {\CCb} {$C^{(2)}_\mathrm{c}$} 
\newcommand {\LCb} {$L^{(2)}_\mathrm{c}$} 
\newcommand {\CMb} {$C^{(2)}_\mathrm{m}$} 
\newcommand {\LMb} {$L^{(2)}_\mathrm{m}$} 

\newcommand {\bCCO} {$\bar{C}_\mathrm{c}$} 
\newcommand {\bLCO} {$\bar{L}_\mathrm{c}$} 
\newcommand {\bCMO} {$\bar{C}_\mathrm{m}$} 
\newcommand {\bLMO} {$\bar{L}_\mathrm{m}$} 
\newcommand {\bCCa} {$\bar{C}^{(1)}_\mathrm{c}$} 
\newcommand {\bLCa} {$\bar{L}^{(1)}_\mathrm{c}$} 
\newcommand {\bCMa} {$\bar{C}^{(1)}_\mathrm{m}$} 
\newcommand {\bLMa} {$\bar{L}^{(1)}_\mathrm{m}$} 
\newcommand {\bCCb} {$\bar{C}^{(2)}_\mathrm{c}$} 
\newcommand {\bLCb} {$\bar{L}^{(2)}_\mathrm{c}$} 
\newcommand {\bCMb} {$\bar{C}^{(2)}_\mathrm{m}$} 
\newcommand {\bLMb} {$\bar{L}^{(2)}_\mathrm{m}$} 

\begin{document}
\title{Unraveling Spurious Properties of Interaction Networks\\with Tailored Random Networks}
\author{Stephan \surname{Bialonski}}
\email{bialonski@gmx.net}
\affiliation{Department of Epileptology, University of Bonn, Bonn, Germany} 
\affiliation{Helmholtz Institute for Radiation and Nuclear Physics, University of Bonn, Bonn, Germany}
\affiliation{Interdisciplinary Center for Complex Systems, University of Bonn, Bonn, Germany} 
\author{Martin \surname{Wendler}}
\affiliation{Fakult\"at f\"ur Mathematik, Ruhr-Universit\"at Bochum, Bochum, Germany} 
\author{Klaus \surname{Lehnertz}}
\affiliation{Department of Epileptology, University of Bonn, Bonn, Germany} 
\affiliation{Helmholtz Institute for Radiation and Nuclear Physics, University of Bonn, Bonn, Germany}
\affiliation{Interdisciplinary Center for Complex Systems, University of Bonn, Bonn, Germany} 

\received{June 1, 2011}
\accepted{July 2, 2011}
\published{August 5, 2011}

\begin{abstract}
We investigate interaction networks that we derive from multivariate time series with methods frequently employed in diverse scientific fields such as biology, quantitative finance, physics, earth and climate sciences, and the neurosciences. Mimicking experimental situations, we generate time series with finite length and varying frequency content but from independent stochastic processes. Using the correlation coefficient and the maximum cross-correlation, we estimate interdependencies between these time series. With clustering coefficient and average shortest path length, we observe unweighted interaction networks, derived via thresholding the values of interdependence, to possess non-trivial topologies as compared to \ER{} networks, which would indicate small-world characteristics. These topologies reflect the mostly unavoidable finiteness of the data, which limits the reliability of typically used estimators of signal interdependence. We propose random networks that are tailored to the way interaction networks are derived from empirical data. Through an exemplary investigation of multichannel electroencephalographic recordings of epileptic seizures---known for their complex spatial and temporal dynamics---we show that such random networks help to distinguish network properties of interdependence structures related to seizure dynamics from those spuriously induced by the applied methods of analysis.
\end{abstract}
\maketitle

\begin{textblock*}{15cm}(3cm,26cm)
\noindent
Published as: S. Bialonski, M. Wendler, and K. Lehnertz. Unraveling spurious properties  of interaction networks with tailored random networks. PLoS ONE, 6(8):e22826, 2011.\\
\noindent DOI: \href{http://dx.doi.org/10.1371/journal.pone.0022826}{10.1371/journal.pone.0022826}
\end{textblock*}

\section*{Introduction}
\enlargethispage{\baselineskip}

The last years have seen an extraordinary success of network theory and its applications in diverse disciplines, ranging from sociology, biology, earth and climate sciences, quantitative finance, to physics and the neurosciences \cite{Newman2003,Boccaletti2006a,Arenas2008,BarratBook2008}. There is now growing evidence that research into the dynamics of complex systems profits from a network perspective. Within this framework, complex systems are considered to be composed of interacting subsystems. This view has been adopted in a large number of modeling studies and empirical studies. It is usually assumed that the complex system under study can be described by an \emph{interaction network}, whose nodes represent subsystems and whose links represent interactions between them. Interaction networks derived from empirical data (multivariate time series) have been repeatedly studied in climate science (climate networks, see \cite{Tsonis2004,Yamasaki2008,Donges2009,Tsonis2010,Steinhaeuser2011} and references therein), in seismology (earthquake networks, see, e.g., \cite{Abe2004,Abe2006,Jimenez2008,Mohan2011}), in quantitative finance (financial networks, see e.g. \cite{Mantegna1999,Onnela2004,Boginski2005,Qiu2010,Emmert-Streib2010} and references therein), and in the neurosciences (brain functional networks, see \cite{Reijneveld2007,Bullmore2009} for an overview). Many interaction networks have been reported to possess non-trivial properties such as small-world architectures, community structures, or hubs (nodes with high centrality), all of which have been considered to be characteristics of the dynamics of the complex system.

When analyzing empirical data one is faced with the challenge of defining nodes and inferring links from multivariate noisy time series with only a limited number of data points due to stationarity requirements. Different approaches varying to some degree across disciplines have been proposed. For most approaches, each single time series is associated with a node and inference of links is based on time series analysis techniques. Bivariate time series analysis methods, such as the correlation coefficient, are used as estimators of signal interdependence which is assumed to be indicative of an interaction between different subsystems. Inferring links from estimates of signal interdependence can be achieved in different ways. Weighted interaction networks can be derived by considering estimated values of signal interdependence (sometimes mapped via some function) as link weights. Since methods characterizing unweighted networks are well-established and readily available, such networks are more frequently derived from empirical data. Besides approaches based on constructing minimum spanning trees (see, e.g., reference \cite{Mantegna1999}), on significance testing \cite{Kramer2009,Donges2009b,Emmert-Streib2010b}, or on rank-ordered network growth (see, e.g., reference \cite{Onnela2004}), a common practice 
pursued in many disciplines is to choose a threshold above which an estimated value of signal interdependence is converted into a link (``thresholding'', see, e.g., references \cite{Tsonis2004,Boginski2005,Jimenez2008,Bullmore2009}). Following this approach, the resulting unweighted interaction networks have been repeatedly investigated employing various networks characteristics, among which we mention the widely-used clustering coefficient $C$ and average shortest path length $L$ to assess a potential small-world characteristic, and the node degrees in order to identify hubs.

As studies employing the network approach grow in numbers, the question arises as to how informative reported results are with respect to the investigated dynamical systems. To address this issue, properties of interaction networks are typically compared to those obtained from network null models. Most frequently, \ER{} random networks \cite{Erdos1959} or random networks with a predefined degree distribution \cite{Rao1996,Maslov2002} serve as null models; network properties
that deviate from those obtained from the null model are considered to be characteristic of the investigated dynamical system. Only in a few recent studies, results obtained from network analyses have been questioned in relation to various assumptions underlying the network analysis approach. Problems pointed out include: incomplete data sets and observational errors in animal social network studies \cite{James2009}; representation issues and questionable use of statistics in biological networks (see \cite{Lima-Mendez2009} and references therein); challenging node and link identification in the neurosciences \cite{Ioannides2007,Butts2009,Bialonski2010}; the issue of spatial sampling of complex systems \cite{Bialonski2010,Antiqueira2010,Gerhard2011}. This calls not only for a careful interpretation of results but also for the development of appropriate null models that incorporate knowledge about the way networks are derived from empirical data.

We study -- from the perspective of field data analysis -- a fundamental assumption underlying the network approach, namely that the multivariate time series are obtained from interacting dynamical processes and are thus well represented by a model of mutual relationships (i.e., an interaction network). Visual inspection of empirical time series typically reveals a perplexing variety of characteristics ranging from fluctuations on different time scales to quasi-periodicity suggestive of different types of dynamics. Moreover, empirical time series are inevitably noisy and finite leading to a limited reliability of estimators of signal interdependencies. This is aggravated with the advent of time-resolved network analyses where a good temporal resolution often comes at the cost of diminished statistics. Taken together, it is not surprising that the suitability of the network approach is notoriously difficult to judge prior to analysis.

We here employ the above-mentioned thresholding-approach to construct interaction networks for which we estimate signal interdependence with the frequently used correlation coefficient and the maximum cross correlation. We derive these
networks, however, from multivariate time series of finite length that are generated by independent (non-interacting) processes which would a priori not advocate the notion of a representation by a model of mutual relationships. In simulation studies we investigate often used network properties (clustering coefficient, average shortest path length, number of connected components). We observe that network properties can deviate pronouncedly from those observed in \ER{} networks depending on the length and the spectral content of the multivariate time series. We address the question whether similar dependencies can also be observed in empirical data by investigating multichannel electroencephalographic (EEG) recordings of epileptic seizures that are known for their complex spatial and temporal dynamics. Finally, we propose random networks that are tailored to the way interaction networks are derived from multivariate empirical time series.

\section*{Methods}
\label{sec:methods}
Interaction networks are typically derived from $N$ multivariate time series $x_i$ ($i\in \{1,\ldots,N\}$) in two steps. First, by employing some bivariate time series analysis method, interdependence between two time series $x_i(t)$ and $x_j(t)$ ($t \in \{1,\ldots,T\}$) is estimated as an indicator for the strength of interaction between the underlying systems. A multitude of estimators \cite{Brillinger1981,Pikovsky_Book2001,Boccaletti2002,Kantz2003,Pereda2005,Hlavackova2007,Lehnertz2009b}, which differ in concepts, robustness (e.g., against noise contaminations), and statistical efficiency (i.e., the amount of data required), is available. Studies that aim at deriving interaction networks from field data frequently employ the absolute value of the linear correlation coefficient to estimate interdependence between two time series. The entries of the correlation matrix \bfR{} then read
\begin{eqnarray}
 \label{eq:corrcoeff}
 \rho_{ij}^\mathrm{c} := \left| T^{-1} \sum_{t=1}^{T} (x_i(t)-\bar{x}_i)(x_j(t)-\bar{x}_j)\hat{\sigma}_i^{-1}\hat{\sigma}_j^{-1} \right| \\
=: \left| \mbox{corr}(x_i,x_j) \right| \text{,}\nonumber 
\end{eqnarray}
where $\bar{x}_i$ and $\hat{\sigma}_i$ denote mean value and the estimated standard deviation of time series $x_i$. Another well established method to characterize interdependencies is the cross correlation function. Here we use the maximum value of the absolute cross correlation between two time series,
\begin{equation}
\label{eq:maxcross}
 \rho_{ij}^\mathrm{m} := \max_{\tau} \left\{ \left| \frac{\xi(x_i,x_j)(\tau)}{\sqrt{\xi(x_i,x_i)(0)\xi(x_j,x_j)(0)}} \right| \right\},
\end{equation}
with
\begin{equation}
\xi(x_i,x_j)(\tau) := \begin{cases} \sum_{t=1}^{T-\tau} x_i(t+\tau) x_j(t) & , \tau \geq 0 \\ \xi(x_j,x_i)(-\tau) &, \tau < 0\end{cases}
\end{equation}
to define the entries of the cross correlation matrix \bfRm{}. As practiced in field data analysis, we normalize the time series to zero mean before pursuing subsequent steps of analysis. Note that $\rho^\mathrm{m}_{ij}$ is then the maximum value of the absolute cross covariance function. 
Both interdependence estimators are symmetric ($\rho_{ij}^\mathrm{c}=\rho_{ji}^\mathrm{c}$ and $\rho^\mathrm{m}_{ij}=\rho^\mathrm{m}_{ji}$) and are confined to the interval $[0,1]$. High values indicate strongly interdependent time series while dissimilar time series result in values close to zero for $T$ sufficiently large.

Second, the adjacency matrix $\mathbf{A}$ representing an unweighted undirected interaction network is derived from \bfR{} (or \bfRm{}) by thresholding. For a threshold $\theta \in [0,1]$ entries $A_{ij}$ and $A_{ji}$ of $\mathbf{A}$ are set to $1$ (representing an undirected link between nodes $i$ and $j$) for all entries $\rho^\mathrm{c}_{ij} > \theta$ ($\rho^\mathrm{m}_{ij} > \theta$, respectively) with $i\neq j$, and to zero (no link) otherwise. In many studies $\theta$ is not chosen directly but determined such that the derived network possesses a previously specified mean degree $\bar{k}:=N^{-1}\sum_i k_i$, where $k_i$ denotes the degree of $i$, i.e., the number of links connected to node $i$. More frequently, $\theta$ is chosen such that the network possesses a previously specified link density $\epsilon = \bar{k}(N-1)^{-1}$. We will follow the latter approach and derive networks for fixed values of $\epsilon$.

To characterize a network as defined by $\mathbf{A}$, a plethora of methods have been developed. Among them, the clustering coefficient $C$ and the average shortest path length $L$ are frequently used in field studies. The local clustering coefficient $C_i$ is defined as 
\begin{equation}
 C_i := \left\{ \begin{array}{cl} \frac{1}{k_i (k_i-1)} \sum_{j,m} A_{ij} A_{jm} A_{mi}, & \mbox{if }k_i > 1\\ 0, & \mbox{if } k_i \in \{0,1\}\mbox{.}\end{array}\right.
\end{equation} 
$C_i$ represents the fraction of the number of existing links between neighbors of node $i$ among all possible links between these neighbors \cite{Watts1998,Newman2003,Boccaletti2006a}. The clustering coefficient $C$ of the network is defined as the mean of the local clustering coefficients,
\begin{equation}
 C := \frac{1}{N} \sum_{i=1}^N C_i\mbox{.}
\end{equation}
$C$ quantifies the local interconnectedness of the network and $C_i,C\in [0,1]$.

The average shortest path length is defined as the average shortest distance between any two nodes,
\begin{equation}\label{eq:L1}
 \tilde{L} := \frac{2}{N(N+1)} \sum_{i\leq j} l_{ij}\mbox{,}
\end{equation}
and characterizes the overall connectedness of the network. $l_{ij}$ denotes the length of the shortest path between nodes $i$ and $j$. The definition of the average shortest path length varies across the literature. Like some authors, we here include the distance from each node to itself in the average ($l_{ii}=0 \forall i$). Exclusion will, however, just change the value by a constant factor of $(N+1)/(N-1)$. 

If a network disintegrates into a number $N_c$ of different connected components, there will be pairs of nodes $(i,j)$, for which no connecting path exists, in which case one usually sets $l_{ij} = \infty$ and thus $\tilde{L}=\infty$. In order to avoid this situation, in some studies $l_{ij}$ in equation (\ref{eq:L1}) is replaced by $l^{-1}_{ij}$. The quantity defined this way is called efficiency \cite{Latora2001,Latora2003}. Another approach, which we will follow here and which is frequently used in field studies, is to exclude infinite values of $l_{ij}$ from the average. The average shortest path length then reads
\begin{equation}
 L := \frac{1}{|S|} \sum_{(i,j) \in S} l_{ij}\mbox{,}
\label{eq:L2}
\end{equation}
where 
\begin{equation}
 S := \{(i,j) \mid l_{ij} < \infty;\mbox{ } i,j = 1,\ldots,N \}
\end{equation}
denotes the set of all pairs $(i,j)$ of nodes with finite shortest path. The number of such pairs is given by $|S|$. Note that $L \rightarrow 0$ for $N_c \rightarrow N$. 

In field studies, values of $C$ and $L$ obtained for interaction networks are typically compared with average values obtained from an ensemble of random \ER{} (ER) networks \cite{Erdos1959}. 
Between every pair of nodes is a link with probability $\epsilon$, and links for different pairs exist independently from each other. 
The expectation value of the clustering coefficient of ER networks is $C_\mathrm{ER} = \epsilon$ \cite{Boccaletti2006a}. 
The dependence of the average shortest path length $L_\mathrm{ER}$ of ER networks on $\epsilon$ and $N$ is more complicated (see references  \cite{Chung2001,Boccaletti2006a}).
Almost all ER networks are connected, if $\epsilon \gg \ln{N}/(N-1)$.
ER networks with a predefined number of links (and thus link density) can also be generated by successively adding links between randomly chosen pairs of nodes until the predefined number of links is reached. During this process, multiple links between nodes are avoided.

\section*{Results}
\subsection*{Simulation studies}

We consider time series $z_i$, $i\in\{1,\ldots,N\}$, whose entries $z_i(t)$
are drawn independently from the uniform probability distribution $\mathcal{U}$ on the interval $(0,1)$. We will later study the impact of different lengths $T$ of these random time series on network properties. In order to enable us to study the effects of different spectral contents on network properties, we add the possibility to low-pass filter $z_i$ by considering
\begin{equation}
 x_{i,M,T}(t) := M^{-1} \sum_{l=t}^{t+M-1} z_i(l),\qquad z_i(l) \sim \mathcal{U}\mbox{,}
\end{equation}
where $t \in \{1,\ldots,T\}$, and $1 \leq M \ll T$. 
By definition $x_{i,1,T}(t) = z_i(t)\forall t$. With the size $M$ of the moving average we control the spectral contents of time series.
We here chose this ansatz for the sake of simplicity, for its mathematical treatability, and because the random time series with different spectral contents produced this way show all properties we want to illustrate.

In the following we will study the influence of the length $T$ of time series on network properties by considering $x_{i,1,T}$ for different $T$. For a chosen value of $T$ we determine $R$ realizations of $x_{i,1,T}$ and we denote each realization $r$ with $x^{(r)}_{i,1,T}$.  When studying the influence of the spectral content we will consider $x_{i,M,T^\prime}$ with different $M$ and with $T^\prime = 500$. We chose this value of $T^\prime$ because we are interested in investigating time series of short length as typically considered in field studies. For a chosen value of $M$ we determine $R$ realizations of $x_{i,M,T^\prime}$ and we denote realization $r$ with $x^{(r)}_{i,M,T^\prime}$.

In order to keep the line of reasoning short and clear, we will present supporting and more rigorous mathematical results in Appendix A and refer to them in places where needed. In addition, since we observed most simulation studies based on \bfRm{} to yield qualitatively the same results as those based on \bfR{}, we will present results based on \bfR{} only and report results of our studies based on \bfRm{} whenever we observed qualitative differences.

\subsubsection*{Clustering coefficient}

\begin{figure*}[tbh]
\includegraphics[width=0.9\textwidth]{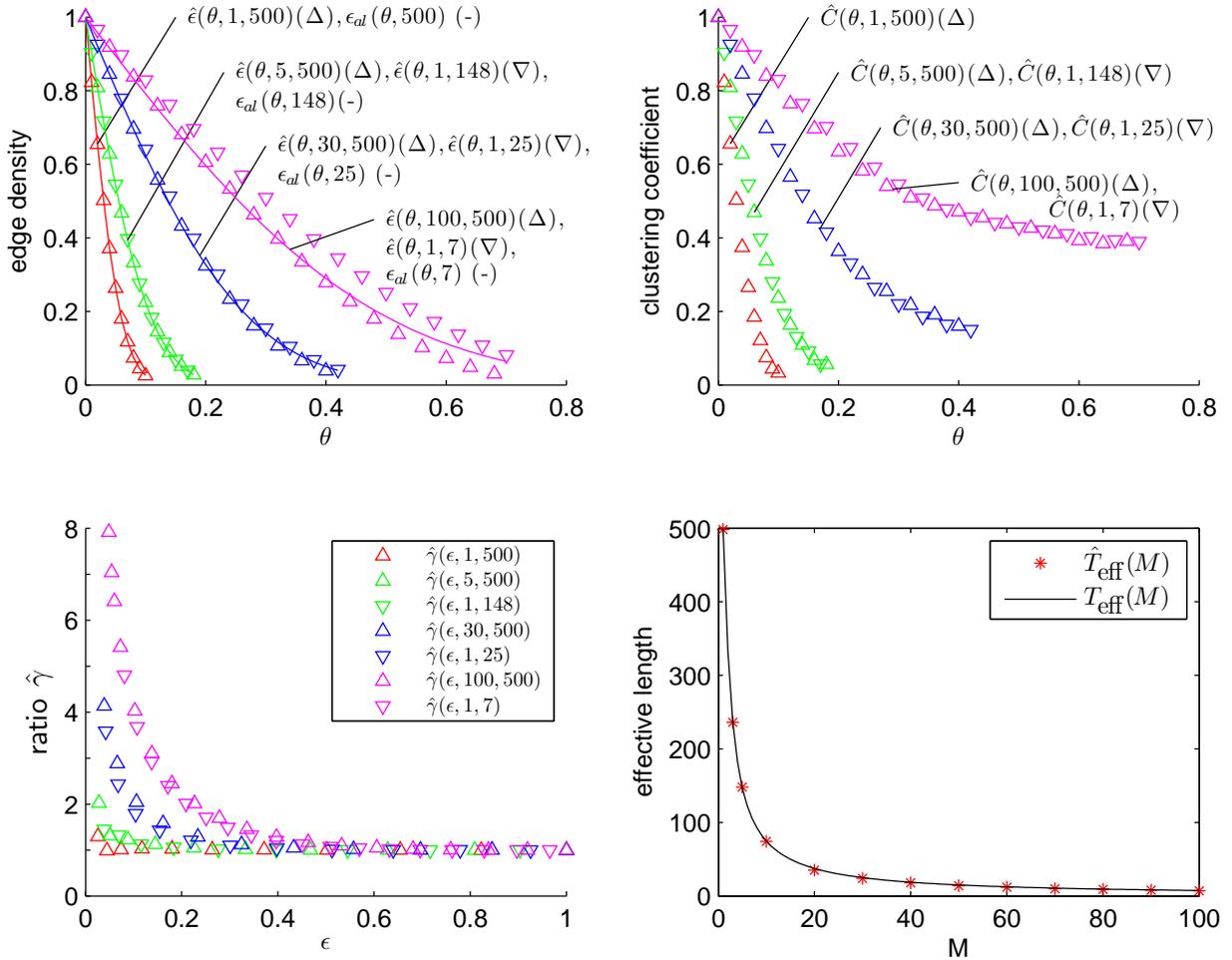}
\caption{{\bf Simulation results for the edge density, the clustering coefficient, and the effective length.} Top row: Dependence of edge density $\hat{\epsilon}(\theta,M,T)$ (left) and of clustering coefficient $\hat{C}(\theta,M,T)$ (right) on the threshold $\theta$ for different values of the size $M$ of the moving average and of the length $T$ of time series. Values of edge density $\epsilon_\mathrm{al}(\theta,T)$ obtained with the asymptotic limit (equation (\ref{eq:epsilon})) are shown as lines (top left). Bottom left: Dependence of the ratio $\hat{\gamma}(\epsilon,M,T)=\hat{C}_{M,T}(\epsilon)/C_\mathrm{ER}(\epsilon)$ on edge density $\epsilon$. Note, that we omitted values of estimated quantities obtained for $\theta \in \{ \theta : (R^{-1}\sum_r H_{12,M,T}^{(r)}(\theta)H_{13,M,T}^{(r)}(\theta)) < 10^{-3} \}$ since the accuracy of the statistics is no longer guaranteed. Bottom right: Dependence of effective length $T_\mathrm{eff}$ as determined by equation (\ref{eq:Teff}) (black line) and its numerical estimate $\hat{T}_\mathrm{eff}$ (red markers) on $M$.}
\label{fig:01}
\end{figure*}

Let $\rho^{(r)}_{ij,1,T} := \rho^\mathrm{c}(x^{(r)}_{i,1,T},x^{(r)}_{j,1,T})$ denote the absolute value of the empirical correlation coefficient estimated for time series $x^{(r)}_{i,1,T}$ and $x^{(r)}_{j,1,T}$, and let us consider $R$ realizations, $r\in\{1,\ldots,R\}$. Because of the independence of processes generating the time series and because of the symmetry of the correlation coefficient, we expect the $R$ values of the empirical correlation coefficient calculated for finite and fixed $T$ to be distributed around the mean value $0$. The variance of this distribution will be higher the lower we choose $T$. If we sample one value $\rho^{(r)}_{ij,1,T}$ out of the $R$ values it is almost surely that $\rho^{(r)}_{ij,1,T}>0$. Thus there are thresholds $0 < \theta < \rho^{(r)}_{ij,1,T}$ for which we would establish a link between the corresponding nodes $i$ and $j$ when deriving a network. Let us now consider a network of $N$ nodes whose links are derived from $N$ time series as before. For some $\theta>0$ the network will possess links and $\epsilon > 0$. We expect to observe $\epsilon$ for some fixed $\theta>0$ to be higher the larger the variance of the distribution of $\rho^{(r)}_{ij,1,T}$.
 Likewise, for fixed values of $\epsilon$ we expect to find $\theta$ to be higher the lower we choose a value of $T$.

As a first check of this intuition we derive an approximation $\epsilon_\mathrm{al}$ for the edge density by making use of the asymptotic limit ($T \rightarrow \infty$, see Appendix A, Lemma 2 for details),
\begin{equation}\label{eq:epsilon}
 \epsilon_\mathrm{al}(\theta,T) = 2 \Phi(-\sqrt{T}\theta)\mbox{,}
\end{equation}
where $\Phi$ denotes the cumulative distribution function of a standard normal distribution. In figure~\ref{fig:01} (top left) we show the dependence of $\epsilon_\mathrm{al}(\theta,T)$ on $\theta$ for exemplary values of $T$. Indeed, $\epsilon_\mathrm{al}(\theta,T)$ is decreasing in $\theta$ and $T$.

The concession of taking the asymptotic limit when deriving equation (\ref{eq:epsilon}) may limit its validity in the case of small values of $T$ in which we are especially interested. 
Thus, we approach this case by simulation studies. Let us consider $R=10^6$ values of $\rho_{12,M,T}^{(r)}$ obtained for $R$ realizations of two time series $x_{i,M,T}$, $i \in \{1,2\}, r \in \{1,\ldots,R\}$. We estimate the edge density $\hat{\epsilon}(\theta,M,T)$ by
\begin{equation}\label{eq:epsmodel}
\hat{\epsilon}(\theta,M,T) := R^{-1} \sum_r H_{12,M,T}^{(r)}(\theta)\mbox{,}
\end{equation}
where $H_{ij,M,T}^{(r)}(\theta) = 1$ for $\rho_{ij,M,T}^{(r)} > \theta$, and $0$ else. Note that $\hat{\epsilon}(\theta,M,T)$ does not depend on $N$. This is because $\hat{\epsilon}(\theta,M,T)$ represents the (numerically determined) probability that there is a link between two vertices. The dependence of $\hat{\epsilon}(\theta,1,T)$ on $\theta$ for different values of $T$ shown in figure~\ref{fig:01} (top left) indicates a good agreement between $\epsilon_\mathrm{al}(\theta,T)$ and $\hat{\epsilon}(\theta,1,T)$ for larger values of $T$ but an increasing difference for $T<30$.

We proceed by estimating the clustering coefficient $\hat{C}$ for our model using $R$ realizations of three time series $x_{i,M,T}$, $i\in \{1,\ldots,3\}$  by
\begin{equation}
\label{eq:clustcoeff}
\hat{C}(\theta,M,T) := \frac{\sum_r H_{12,M,T}^{(r)}(\theta)H_{13,M,T}^{(r)}(\theta)H_{23,M,T}^{(r)}(\theta) } {\sum_r H_{12,M,T}^{(r)}(\theta)H_{13,M,T}^{(r)}(\theta)}\mbox{.}
\end{equation}
The dependence of $\hat{C}(\theta,1,T)$ on $\theta$ for various $T$ is shown in the top right part of figure~\ref{fig:01}.
For fixed $T$, $\hat{C}(\theta,1,T)$ decreases from $1$ with increasing values of $\theta$ which one might expect due to the decrease of $\epsilon$. However, we also observe for $\theta > 0$ that $\hat{C}(\theta,1,T)$ takes on higher values the lower $T$.

In order to investigate whether the clustering coefficients of our networks differ from those of Erd\H{o}s-R\'{e}nyi networks we use equation (\ref{eq:epsmodel}) and obtain $\hat{C}_{M,T}(\epsilon) = \hat{C}(\hat\theta(\epsilon,M,T),M,T)$ with $\hat{\theta}(\epsilon,M,T)=\inf\{\theta:\ \hat{\epsilon}(\theta,M,T)\geq \epsilon\}$. This allows the comparison with  $C_\mathrm{ER}(\epsilon)=\epsilon$ by considering the ratio $\hat{\gamma}(\epsilon,M,T):=\hat{C}_{M,T}(\epsilon)/C_\mathrm{ER}(\epsilon)$. Remarkably, $\hat{\gamma}(\epsilon,1,T) \gg 1$ for a range of values of $\epsilon$ and $T$ (cf. lower left part of figure~\ref{fig:01}).
$\hat{\gamma}(\epsilon,1,T)$ even increases for small $\epsilon$ and $T$. This indicates that there is a relevant dependence between the three random variables $\rho_{ij,M,T}$, $\rho_{il,M,T}$, and $\rho_{jl,M,T}$ for different indices $i,j,l$ and small $T$. For $T\rightarrow\infty$ and fixed edge density, $C$ converges to $C_\mathrm{ER}$
because the dependence between the random variables $\rho_{ij,M,T}$, $i,j\in\{1,\ldots,N\}$, vanishes (i.e., the random variables will converge in distribution to independent normal random variables).\\ 

In order to gain deeper insights into the influence of the spectral contents of random time series on the clustering coefficient, we repeat the steps of analysis with time series $x_{i,M,T^\prime}$, where $T^\prime=500$ is kept fix, and we choose different values of $M$. Figure \ref{fig:01} (top panels and lower left) shows that the higher the amount of low-frequency contributions (large $M$)
the higher
$\hat{\epsilon}(\theta,M,T^\prime)$ and $\hat{C}(\theta,M,T^\prime)$ (for $\theta>0$), and the higher $\hat{\gamma}(\epsilon,M,T^\prime)$ (for $\epsilon \ll 1$). The difference between Erd\H{o}s-R\'{e}nyi networks and our time series networks becomes more pronounced ($\hat{\gamma}(\epsilon,M,T^\prime)\gg 1$) the smaller $\epsilon$ and the higher $M$.\\

\noindent
Given the similar dependence of $\hat{\gamma}$, $\hat{C}$, and $\hat{\epsilon}$ on $T$ and $M$, 
we hypothesize that the similarity can be traced back to similar variances of $\rho_{ij,1,T}$ and $\rho_{ij,M,T^\prime}$ for some values of $T$ and $M$. 
By making use of the asymptotic variance of the limit distributions of $T\rightarrow \infty$, we derive an expression relating Var$(\rho_{ij,1,T})$ and Var$(\rho_{ij,M,T})$ to each other (see Appendix A, Lemma 1),
\begin{multline}\label{eq:variances}
 \mathrm{Var}(\rho_{ij,M,T}) \approx g(M) \mathrm{Var}(\rho_{ij,1,T})\text{,}\\\text{ with }g(M) = \frac{2}{3}M+\frac{1}{3M}\mbox{.}
\end{multline}
We are now able to define an effective length $T_\mathrm{eff}$ of time series,
\begin{equation}\label{eq:Teff}
 T_\mathrm{eff}(M) := \frac{T^\prime}{g(M)}\mbox{,}
\end{equation}
for which $\mbox{Var}(\rho_{ij,1,T_\mathrm{eff}}) \approx \mbox{Var}(\rho_{ij,M,T^\prime})$.
In the lower right part of figure~\ref{fig:01} we show $T_\mathrm{eff}(M)$ in dependence on $M$. 
To investigate whether equation (\ref{eq:Teff}) also holds for small values of $T$,
we determine numerically, for different values of $\theta$, $\hat{C}(\theta,1,T)$ for $T \in \{3,\ldots,T^\prime\}$ as well as $\hat{C}(\theta,M,T^\prime)$ for some chosen values of $M$. Eventually, we determine 
for each value of $M$ a value of $T$, for which $\hat{C}(\theta,1,T)$ and $\hat{C}(\theta,M,T^\prime)$ curves match in a least-squares sense, and denote this value with $\hat{T}_\mathrm{eff}$ (see the lower right part of figure~\ref{fig:01}).
We observe a maximum deviation $\left| T_\mathrm{eff}-\hat{T}_\mathrm{eff} \right|\approx 2$ and conclude that equation (\ref{eq:Teff}) indeed holds for small length $T$ of time series.
Moreover, numerically determined dependencies of $\hat{\epsilon}$ on $\theta$, $\hat{C}$ on $\theta$, as well as $\hat{\gamma}$ on $\epsilon$ for pairs of values $(M,T^\prime)$ show a remarkable similarity to those dependencies obtained for pairs of values $(1,\hat{T}_\mathrm{eff})$.

\begin{figure*}[bth]
\includegraphics[width=0.85\textwidth]{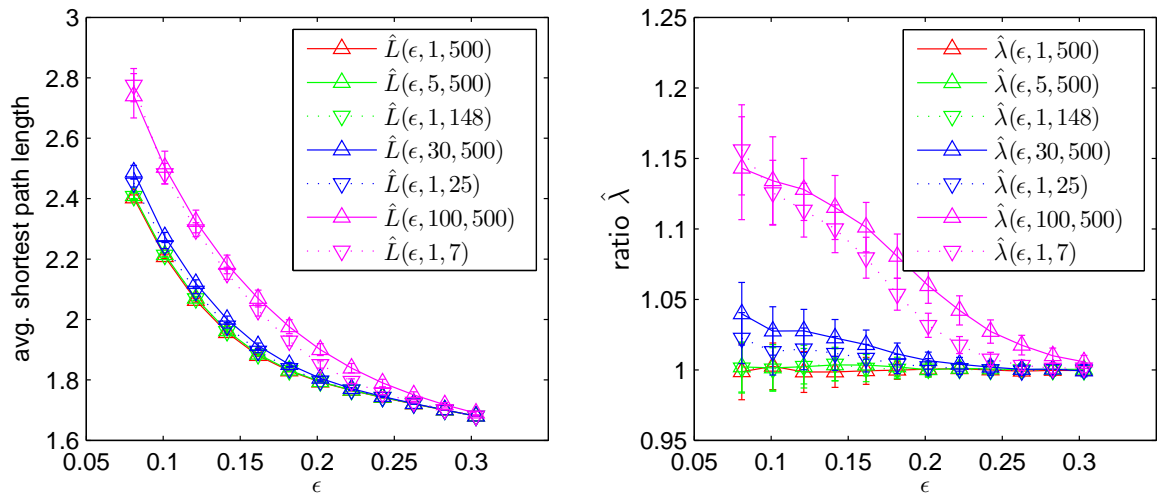}
\caption{{\bf Simulation results for the average shortest path length.} Dependence of the average shortest path length $\hat{L}(\epsilon,M,T)$ (left) and of the ratio $\hat{\lambda}(\epsilon,M,T)=\hat{L}(\epsilon,1,T)/L_\mathrm{ER}(\epsilon)$ (right) on edge density $\epsilon$ for different values of the size $M$ of the moving average and of the length $T$ of time series. Lines are for eye-guidance only.}
\label{fig:02}
\end{figure*}

Thus, the clustering coefficient of networks derived from random time series of finite length and/or with a large amount of low-frequency contributions is higher than the one of Erd\H{o}s-R\'{e}nyi (ER) networks -- independently of the network size $N$ (cf. equation (\ref{eq:clustcoeff})). This difference becomes more pronounced the lower the edge density $\epsilon$, the lower the length $T$ of time series, and the larger the amount of low-frequency contributions. These results point us to an important difference between ER networks and our model networks: possible edges in ER networks are not only (1) equally likely but also (2) independently chosen to become edges. While property (1) is fulfilled in our model networks, property (2) is not.

\subsubsection*{Average shortest path length}

Next we study the impact of the length of time series and of the amount of low-frequency contributions
on the average shortest path length $L$ of our model networks by employing a similar but different simulation approach as used in the previous section. To estimate $L$, we consider $R=100$ networks with a fixed number of nodes ($N=100$). We derive our model networks by thresholding $\rho^{(r)}_{ij,M,T}$, $i,j\in\{1,\ldots,N\}$, $r\in \{1,\ldots,R\}$. Let $L^{(r)}(\epsilon,1,T)$ denote the average shortest path length for network $r$ with $M=1$ and different values of $T$, and $L^{(r)}(\epsilon,M,T^\prime)$ the average shortest path length for network $r$ with fixed value of $T$ ($T=T^\prime=500$) and different values of $M$. With $L^{(r)}_\mathrm{ER}(\epsilon)$ we refer to the average shortest path length obtained for the $r$-th ER network of size $N$ and edge density $\epsilon$. Mean values over realizations will be denoted as $\hat{L}(\epsilon,1,T)$, $\hat{L}(\epsilon,M,T^\prime)$, and $\hat{L}_\mathrm{ER}(\epsilon)$ respectively. Finally, we define $\hat{\lambda}(\epsilon,1,T) := \hat{L}(\epsilon,1,T)/\hat{L}_\mathrm{ER}(\epsilon)$ and $\hat{\lambda}(\epsilon,M,T^\prime) := \hat{L}(\epsilon,M,T^\prime)/\hat{L}_\mathrm{ER}(\epsilon)$.

In figure~\ref{fig:02} we show the dependence of $\hat{L}$
and $\hat{\lambda}$ on $\epsilon$ for various values of $M$ and $T$. All quantities decrease as $\epsilon$ increases which can be expected due to additional edges reducing the average distances between pairs of nodes of the networks. For fixed $\epsilon \ll 1$, $\hat{L}$ takes on higher values the higher $M$ or the lower $T$. With equation (\ref{eq:Teff}) we have $\hat{L}(\epsilon,1,T_\mathrm{eff}) \approx \hat{L}(\epsilon,M,T^\prime)$ which resembles the results obtained for the clustering coefficient. Differences between the average shortest path lengths of our model networks and ER networks (as characterized by $\hat{\lambda}$) become more pronounced the higher $M$ and the lower $T$. For edge densities typically reported in field studies ($\epsilon\approx 0.1$), however, differences are less pronounced ($\hat{\lambda} \lesssim 1.2$, cf. figure~\ref{fig:02} right) than the ones observed for the clustering coefficient ($\hat{\gamma}>2$ for selected values of $M$ and $T$, cf. figure~\ref{fig:01} bottom left). We obtained qualitatively similar results for small ($N=50$) and large numbers of nodes ($N=500$).

\subsubsection*{Number of connected components and degree distribution}

\begin{figure*}[tbh]
\includegraphics[width=0.8\textwidth]{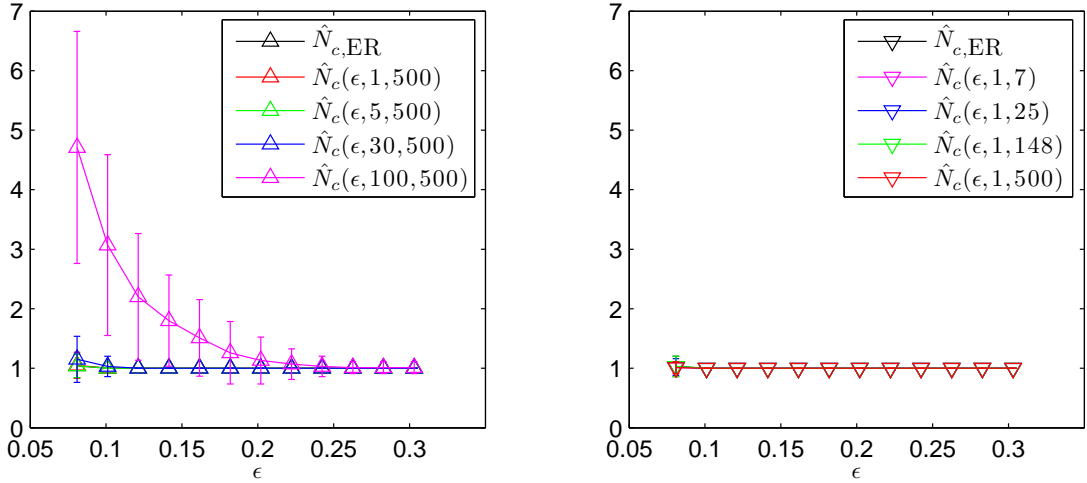}
\caption{{\bf Simulation results for the number of connected components.} Dependence of the number of connected components $\hat{N}_c(\epsilon,M,T)$ on the edge density $\epsilon$ for different values of the size $M$ of the moving average (left, for $T=500$) and of the length $T$ of time series (right, for $M=1$). Lines are for eye-guidance only.}
\label{fig:02B}
\end{figure*}

\begin{figure*}[tbh]
\includegraphics[width=0.8\textwidth]{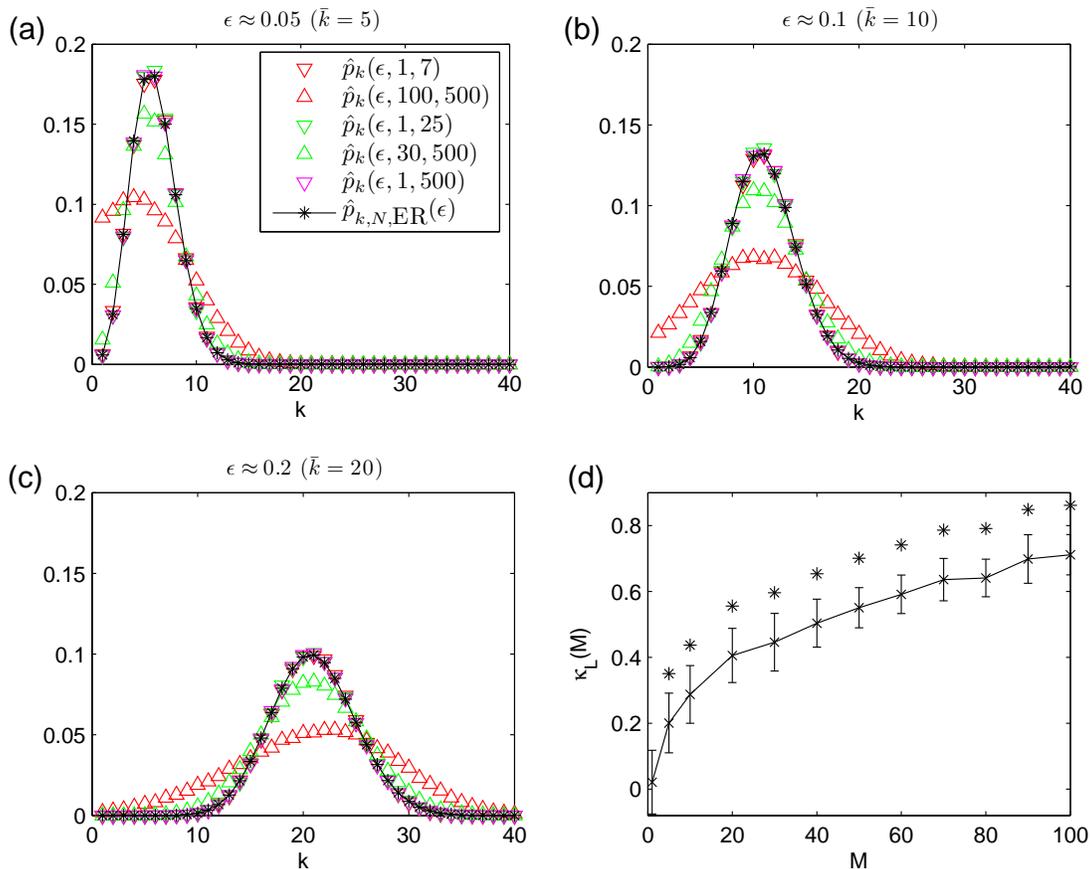}
\caption{{\bf Simulation results for the degree distribution.} ({\bf a-c}) Degree distributions $\hat{p}_k(\epsilon,M,T)$ estimated for $R=1000$ realizations of networks derived from time series $x_{i,M,T}$ ($N=100$) via thresholding using various edge densities $\epsilon=\bar{k}(N-1)^{-1}$ and for selected values of the size $M$ of the moving average and of the length $T$ of time series. The symbol legend in (a) also holds for (b) and (c). ({\bf d}) Dependence of correlation ($\kappa_L(M)$) between node degrees and power content in the lower frequency range on the size $M$ of the moving average. Mean values of correlations obtained for $R=100$ realizations of networks for each value of $M$ are shown as crosses and standard deviations as error bars. Stars indicate significant differences in comparison to $\kappa_L(1)$ (Bonferroni corrected pair-wise Wilcoxon rank sum tests for equal medians, $p<0.01$). Lines are for eye-guidance only.}
\label{fig:03}
\end{figure*}

Since the number of connected components of a given network might affect network characteristics such as the average shortest path length (see equation (\ref{eq:L2})), we investigate the impact of different length of time series and of the amount of low-frequency contributions on the average number of connected components $\hat{N}_c(\epsilon,M,T)$ of the networks derived from $x_{i,1,T_\mathrm{eff}}$ and $x_{i,M,T^\prime}$.
We determine $\hat{N}_c(\epsilon,M,T)$ as the mean of $R$ realizations of the corresponding networks and with
$\hat{N}_{c,\mathrm{ER}}(\epsilon)$ we denote the mean value of the number of connected components in $R$ realization of ER networks. For the edge densities considered here we observe ER networks to be connected (cf. figure~\ref{fig:02B}), $N_{c,\mathrm{ER}} \approx 1$, which is in agreement with the connectivity condition for ER networks, $\epsilon \gg \ln{N}/(N-1) \approx 0.05$ (for $N=100$). Similarly, we observe $\hat{N}_c(\epsilon,1,T_\mathrm{eff}) \approx 1$, even for small values of $T$ (cf. figure~\ref{fig:02B} right). 
In contrast, $\hat{N}_c(\epsilon,M,T^\prime)$ takes on higher values the lower $\epsilon$ and the higher $M$ (cf. figure~\ref{fig:02B} left). In order to achieve a better understanding of these findings, we determine degree probability distributions of our model networks. Let $\hat{p}_k$ denote the estimated probability of a node to possess a degree $k$, i.e., $\hat{p}_k = \#\{i^{(r)} : k_i^{(r)} = k, r \in \{1,\ldots,R\} \}/(NR)$.
With $\hat{p}_k(\epsilon,M,T)$ we will denote the estimated degree distribution for networks derived from $x_{i,M,T}$. We briefly recall that the degree distribution of ER networks $p_{k,N,\mathrm{ER}}$ follows a binomial distribution,
\begin{equation}
 p_{k,N,\mathrm{ER}}(\epsilon) = \binom{N-1}{k} \epsilon^k (1-\epsilon)^{N-k-1}\mbox{,}
\end{equation}
which we show in figure~\ref{fig:03} for $N=100$ and various $\epsilon$ (top panels and lower left panel). In the same figure we present our findings for $\hat{p}_k(\epsilon,M,T)$ for various values of $T=T_\mathrm{eff}$ and $M$. We observe $\hat{p}_k(\epsilon,1,T_\mathrm{eff})$ to be equal to $\hat{p}_{k,N,\mathrm{ER}}(\epsilon)$ within the error to be expected due to the limited sample size used for the estimation. For $\hat{p}_k(\epsilon,M,T^\prime)$, however, we observe striking differences in comparison to the previous degree distributions. In particular, for decreasing $\epsilon$ and higher $M$, the probability of nodes with degree $k=0$ increases, which leads to networks with disconnected single nodes, thereby increasing the number of connected components of the network.

We hypothesize that the observed differences in the number of connected components as well as in the degree distributions
are related to differences in the spectral content of different realizations of time series 
$x^{(r)}_{i,M,T^\prime}$ for $M>1$. In particular, a node $i$ with a low degree $k_i$ might be associated with a time series $x^{(r)}_{i,M,T^\prime}$, which possesses, by chance, a small relative amount of low frequency contributions (or, equivalently, a large relative amount of high frequency contributions).

In order to test this hypothesis, we generate $R$ realizations of $N=100$ time series $x_{i,M,T^\prime}^{(r)}$ and
estimate their periodograms $\hat{P}^{(r)}_{i,M}(f)$ for frequencies $f\in\{0,\ldots,f_\mathrm{Nyq}\}$ using a discrete Fourier transform \cite{Press2002}. $f_\mathrm{Nyq}$ denotes the Nyquist frequency, and periodograms are normalized such that $\sum_f\hat{P}^{(r)}_{i,M}(f) = 1$. From the same time series, we then derive the networks using $\epsilon=0.1$ and determine the degrees $k_i^{(r)}$. For some fixed $f^\prime > 0$ we define the total power above $f^\prime$ (upper frequency range)
as $\hat{P}_{i,M}^{\mathrm{H},(r)}= \sum_{f\prime}^{f_\mathrm{Nyq}} \hat{P}^{(r)}_{i,M}(f)$, and the total power 
below $f^\prime$ (lower frequency range) as $\hat{P}_{i,M}^{\mathrm{L},(r)}= \sum_{f=0}^{f^\prime-1} \hat{P}^{(r)}_{i,M}(f)$. 
For each realization $r$ we estimate the correlation coefficients between the degrees and the corresponding total power contents in upper and lower frequency range, $\kappa^{(r)}_\mathrm{L}=\mbox{corr}(k^{(r)},\hat{P}_{M}^{\mathrm{L},(r)})$ and $\kappa^{(r)}_\mathrm{H}=\mbox{corr}(k^{(r)},\hat{P}_{M}^{\mathrm{H},(r)})$, respectively, and
determine their mean values, $\kappa_\mathrm{L}(M) = R^{-1}\sum_r \kappa^{(r)}_\mathrm{L}$ and $\kappa_\mathrm{H}(M) = R^{-1}\sum_r \kappa^{(r)}_\mathrm{H}$. Note that $\kappa_\mathrm{L}(M) = -\kappa_\mathrm{H}(M)$ by construction. We choose $f^\prime = f^\prime(M)$ such that 40\% of the total power of the filter function associated with the moving average is contained within the frequency range $f\in[0,f^\prime]$.

For increasing $M$ we observe in the lower right panel of figure~\ref{fig:03} the degrees to be increasingly correlated with $\hat{P}_{M}^{\mathrm{L},(r)}$, which corresponds to an anti-correlation of degrees with 
$\hat{P}_{M}^{\mathrm{H},(r)}$. Thus, as hypothesized above, the observed differences in the degree distributions can indeed be related to the differences in the power content of the time series. We mention that the exact choice of $f^\prime$ does not sensitively affect the observed qualitative relationships as long as $0 < f^\prime \ll f_\mathrm{Nyq}$ is fulfilled. 

We briefly summarize the results obtained so far, which indicate a striking difference between networks derived from independent random time series using \bfR{} or \bfRm{} (cf. equations (\ref{eq:corrcoeff}) and (\ref{eq:maxcross}))
and corresponding ER networks. First, we observed the clustering coefficient $C$ and the average shortest path length $L$ of our networks to be higher the lower the length $T$ of the time series (cf. figures \ref{fig:01} and \ref{fig:02}). Second, for some fixed $T$ we observed $C$ and $L$ to be higher the larger the amount of low frequency components (as parametrized by $M$) in the time series. In addition, these contributions led to an increasing number of connected components in our networks and to degree distributions that differed strongly from those of the corresponding ER networks (cf. figures~\ref{fig:02B} and \ref{fig:03}). We mention that $L$ as defined here (cf. equation (\ref{eq:L2})) tends to decrease for networks with an increasing number $N_c$ of connected components, and $L \rightarrow 0$ for $N_c \rightarrow N$. $L$ thus depends non-trivially on the amount of low frequency components in the time series.
Third, for small edge densities $\epsilon$ and for short time series lengths or for a large amount of low frequency components,
the clustering coefficient deviates more strongly from the one of corresponding ER networks ($\hat{\gamma}>2$) than the average shortest path length ($\hat{\lambda} \lesssim 1.2$; cf. figure~\ref{fig:02} right and figure~\ref{fig:01} (bottom left)). Networks derived from independent random time series can thus be classified as small world networks if one uses $\gamma\gg1$ and $\lambda\approx 1$ as practical criterion, which is often employed in various field studies (cf. \cite{Bialonski2010} and references therein).

\subsection*{Field data analysis}

\begin{figure*}[tbh]
\includegraphics[width=0.9\textwidth]{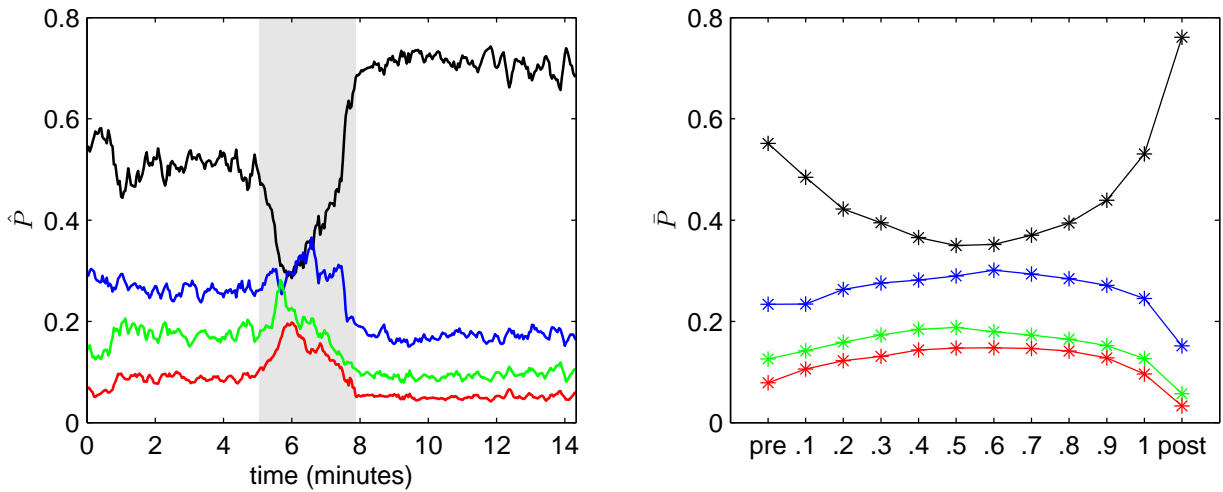}
\caption{{\bf Evolving relative amount of power during epileptic seizures.} (Left) Relative amount of power $\hat{P}$ contained in the $\delta$- ($\hat{P}_\delta$, black), $\vartheta$- ($\hat{P}_\vartheta$, blue), $\alpha$- ($\hat{P}_\alpha$, green), and $\beta$- ($\hat{P}_\beta$, red) frequency bands during an exemplary seizure. Profiles are smoothed using a four-point moving average. Grey-shaded area marks the seizure. (Right) Mean values ($\bar{P}_{\delta}$, $\bar{P}_{\vartheta}$, $\bar{P}_{\alpha}$, $\bar{P}_{\beta}$) of the relative amount of power averaged separately for pre-seizure, discretized seizure, and post-seizure time periods of 100 epileptic seizures. Lines are for eye-guidance only.}
\label{fig:09}
\end{figure*}

The findings obtained in the previous section indicate that strong low frequency contributions affect network properties $C$ and $L$ in a non-trivial way. We now investigate this influence in electroencephalographic (EEG) recordings of epileptic seizures that are known for their complex spatial and temporal changes in frequency content \cite{Franaszczuk1998b,Schiff2000,Jouny2003,Bartolomei2010}. We analyze the multichannel ($N=53\pm 21$ channels) EEGs from 60 patients capturing 100 epileptic seizures reported in reference \cite{Schindler2008a}. All patients had signed informed consent that their clinical data might be used and published for research purposes. The study protocol had previously been approved by the ethics committee of the University of Bonn. During the presurgical evaluation of drug-resistant epilepsy, EEG data were recorded with chronically implanted strip, grid, or depth electrodes from the cortex and from within relevant structures of the brain. The data were sampled at 200 Hz within the frequency band $0.5-70$ Hz using a 16-bit analog-to-digital converter. Electroencephalographic seizure onsets and seizure ends were automatically detected \cite{Schindler2007a}, and EEGs were split into consecutive non-overlapping windows of 2.5 s duration ($T=500$ sampling points). Time series of each window were normalized to zero mean and unit variance. We determined \bfR{} and \bfRm{} for all combinations of EEG time series from each window and derived networks with a fixed edge density $\epsilon = 0.1$ in order to exclude possible edge density effects. With \LCO{} and \CCO{} as well as \LMO{} and \CMO{} we denote characteristics of networks based on \bfR{} and \bfRm{}, respectively. In order to simplify matters, we omit the window indexing in the following.

We investigate a possible influence of the power content of EEG time series on the clustering coefficient and the average shortest path length by comparing their values to those obtained from ensembles of random networks that are based on 
properties of the EEG time series at two different levels of detail. For the first ensemble and for each patient we derive networks from random time series with a power content that approximately equals the mean power content of all EEG time series within a window. Let $\hat{P}_i(f)$ denote the estimated periodogram of each EEG time series $i$, and with $P(f) = N^{-1}\sum_i \hat{P}_i(f)$ we denote the mean power for each frequency component $f$ over all $N$ time series. We normalize $P(f)$ such that $\sum_f P(f) = 1$. We generate $N$ random time series of length $T=500$ whose entries are independently drawn from a uniform probability distribution, and we filter these time series in the Fourier domain using $\sqrt{P(f)}$ as filter function.
We normalize the filtered time series to zero mean and unit variance and derive a network based on \bfR{} or \bfRm{} ($\epsilon=0.1$). We use 20 realizations of such networks per window in order to determine the mean values of network characteristics \CCa{} and \LCa{} as well as \CMa{} and \LMa{} based on \bfR{} or \bfRm{}, respectively. Since the power spectra of all time series equal each other, these random networks resemble the ones investigated in the previous section.

With the second ensemble, we take into account that the power content of EEG time series recorded from different brain regions may differ substantially. For this purpose we make use of a well established method for generating univariate time series surrogates \cite{Schreiber1996a,Schreiber2000a}, which have power spectral contents and amplitude distributions that are practically indistinguishable from those of EEG time series but are otherwise random. Amplitudes are iteratively permuted while the power spectrum of each EEG time series is approximately preserved. Since this randomization scheme destroys any significant linear or non-linear dependencies between time series, it has been successfully applied to test the null hypothesis of independent linear stochastic processes. For each patient, we generated 20 surrogate time series for each EEG time  series from each recording site and each window, and derived networks based on either \bfR{} or \bfRm{} ($\epsilon=0.1$). Mean values of characteristics of these random networks are denoted as \CCb{} and \LCb{} as well as \CMb{} and \LMb{}, respectively.

We begin with an exemplary recording of a seizure of which we show in figure~\ref{fig:09} (left) the temporal evolution of the relative amount of power in the $\delta$- (0--4 Hz, $P_\delta$), $\vartheta$- (4--8 Hz, $P_\vartheta$), $\alpha$- (8--12 Hz, $P_\alpha$), and $\beta$- (12--20 Hz, $P_\beta$) frequency bands. Prior to the seizure the $\delta$-band contains more than 50\% of the total power which is then shifted towards higher frequencies and back towards low frequencies at seizure end. $P_{\delta}$ is even higher after the seizure than prior to the seizure.

\begin{figure*}[tbh]
\includegraphics[width=0.9\textwidth]{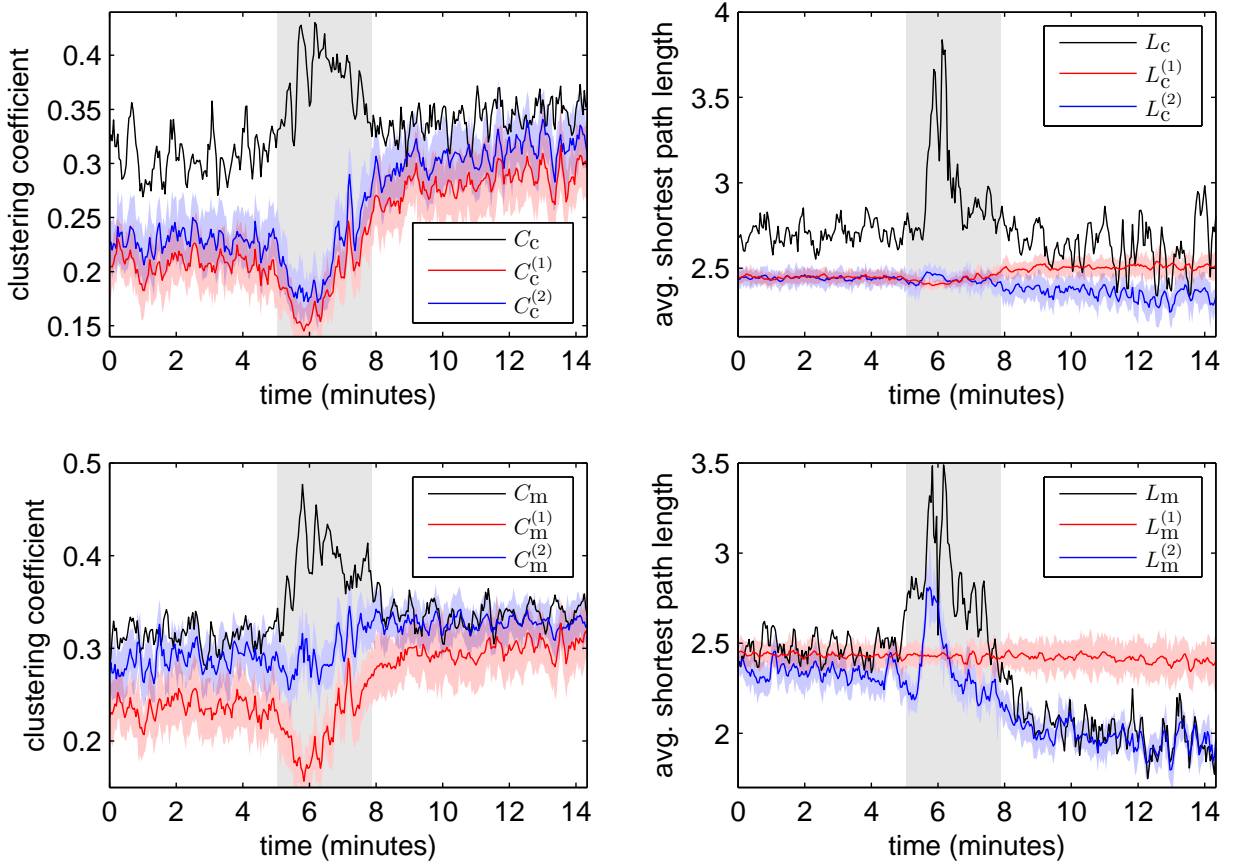}
\caption{{\bf Evolving network properties during an exemplary epileptic seizure.} Network properties \CCO{} and \LCO{} (top row, black lines) as well as \CMO{} and \LMO{} (bottom row, black lines) during an exemplary seizure (cf. figure~\ref{fig:09} (left)). Mean values and standard deviations of network properties obtained from surrogate time series (\CCb{}, \LCb{}, \CMb{}, \LMb{}) are shown as blue lines and blue shaded areas, respectively, and mean values and standard deviations of network properties obtained from the overall power content model (\CCa{}, \LCa{}, \CMa{}, \LMa{}) are shown as red lines and red shaded areas, respectively. Profiles are smoothed using a four-point moving average. Grey-shaded area marks the seizure. For corresponding \ER{} networks $C_\mathrm{ER}\approx 0.1$ and $L_\mathrm{ER}\approx 2.4$ for all time windows.}
\label{fig:05}
\end{figure*}

In figure~\ref{fig:05} we show the temporal evolution of network properties obtained for this recording based on \bfR{} (top panels) and \bfRm{} (bottom panels). During the seizure both the clustering coefficients \CCO{} and \CMO{} and the average shortest path lengths \LCO{} and \LMO{} show pronounced differences to the respective properties obtained from the random networks. These differences are less pronounced prior to and after the seizure, where \CMb{} and \LMb{} even approach the values of \CMO{} and \LMO{}, respectively. \CCa{} and \CMa{} decrease during the seizure and already increase prior to seizure end, resembling the changes of $P_\delta$ (cf. figure~\ref{fig:09} (left)). This is in accordance with results of our simulation studies: there we observed the clustering coefficient to be higher the larger the amount of low frequency components in the time series; this could also be observed, but to a much lesser extent, for the average shortest path length. Indeed, \LMa{} and \LCa{} vary little over time, and \LCa{} is only slightly increased after the seizure, reflecting the high amount of power in the $\delta$-band.

\begin{figure*}[tbh]
\includegraphics[width=0.8\textwidth]{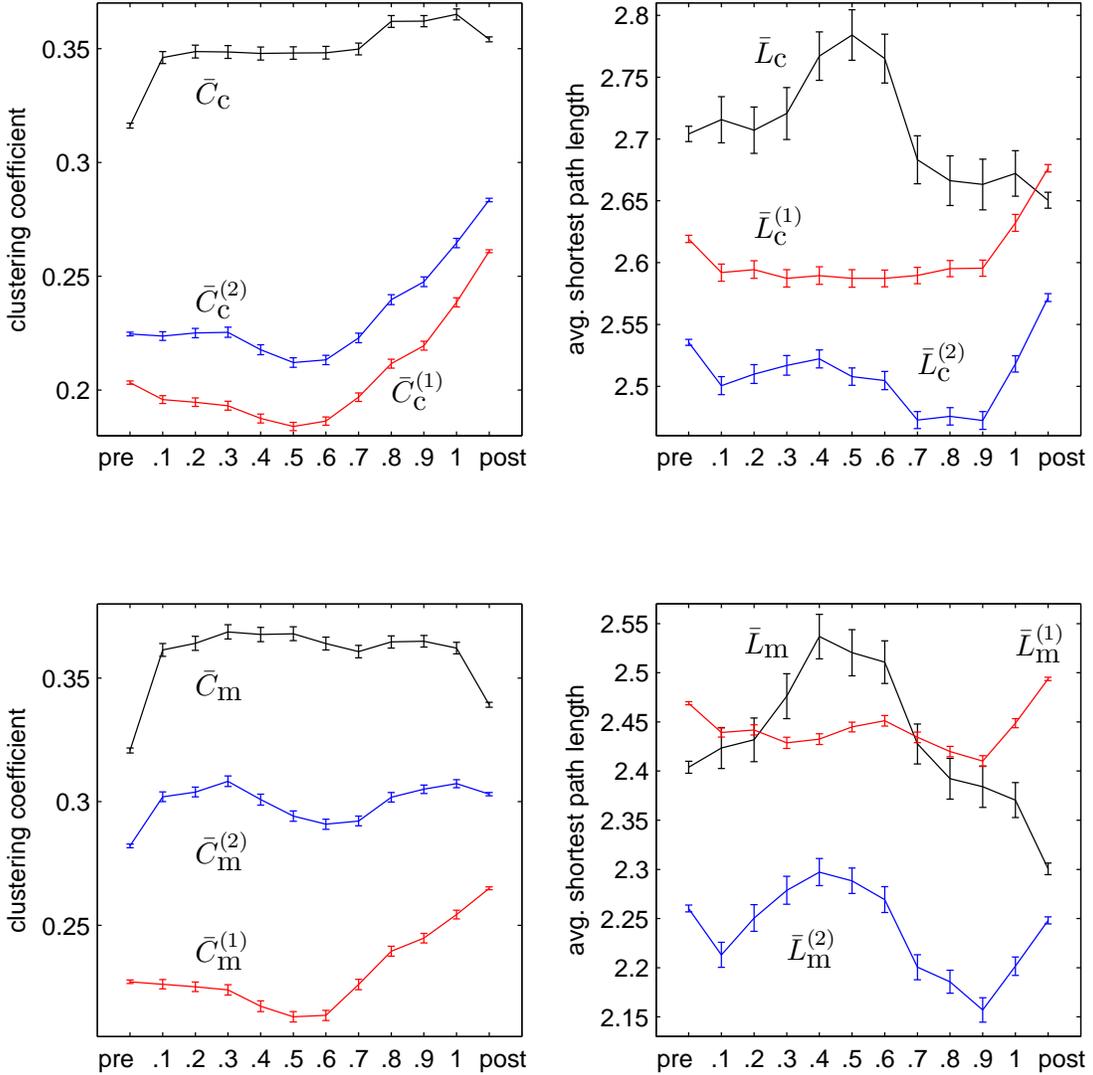}
\caption{{\bf Evolving network properties averaged over 100 epileptic seizures.} Mean values (black) of network properties \CCO{} (top left), \LCO{} (top right), \CMO{} (bottom left), and \LMO{} (bottom right) averaged separately for pre-seizure, discretized seizure, and post-seizure time periods of 100 epileptic seizures. Mean values of corresponding network properties obtained from the first and the second ensemble of random networks are shown as red and blue lines, respectively. All error bars indicate standard error of the mean. Lines are for eye-guidance only.}
\label{fig:07}
\end{figure*}

We only observe small deviations between \CCa{} and \CCb{} as well as between \LCa{} and \LCb{}, which appear to be systematic (for many windows \CCa{}$\lesssim$\CCb{} and \LCa{}$\gtrsim$\LCb{}). These suggest that for interaction networks derived from \bfR{}, both random network ensembles appear appropriate to characterize the influence of power in low frequency bands on clustering coefficient and the average shortest path length. In contrast, we observed differences between \CMa{} and \CMb{}, as well as between \LMa{} and \LMb{}. These differences were most pronounced during the seizure and for \LMa{} and \LMb{} also after the seizure. This finding indicates that the clustering coefficient and average shortest path length of interaction networks derived from \bfRm{} depend sensitively on the power contents of EEG time series recorded from different brain regions. Thus, for these interaction networks only the random networks that account for the complex changes in frequency content of different brain regions prior to, during, and after seizures appear appropriate to characterize the influence of power in low frequency bands on clustering coefficient and the average shortest path length.

We continue by studying properties of networks derived from the EEG recordings of all 100 focal onset seizures. 
Due to the different durations of seizures (mean seizure duration: $110\pm 60$ s) we partitioned each seizure into 10 equidistant time bins (see reference \cite{Schindler2008a} for details) and assigned the time-dependent network properties to the respective time bins. For each seizure we included the same number of pre-seizure and post-seizure windows in our analysis and assigned the corresponding time-dependent network properties to one pre-seizure and one post-seizure time bin. Within each time bin we determined the mean value (e.g., \bCCO{}) and its standard error for each property. In figure~\ref{fig:09} (right), we show for each time bin the mean values of the relative amount of power in different frequency bands of all seizure recordings ($\bar{P}_{\delta}$, $\bar{P}_{\vartheta}$, $\bar{P}_{\alpha}$, $\bar{P}_{\beta}$). Similar to the exemplary recording (cf. figure~\ref{fig:09} (left)), we observe a shift in the relative amount of power in low frequencies prior to seizures towards higher frequencies during seizures and back to low frequencies at seizure end. The amount of power in the $\delta$-band is on average higher in the post-seizure bin than in the pre-seizure bin.

\begin{figure*}[tbh]
\includegraphics[width=0.8\textwidth]{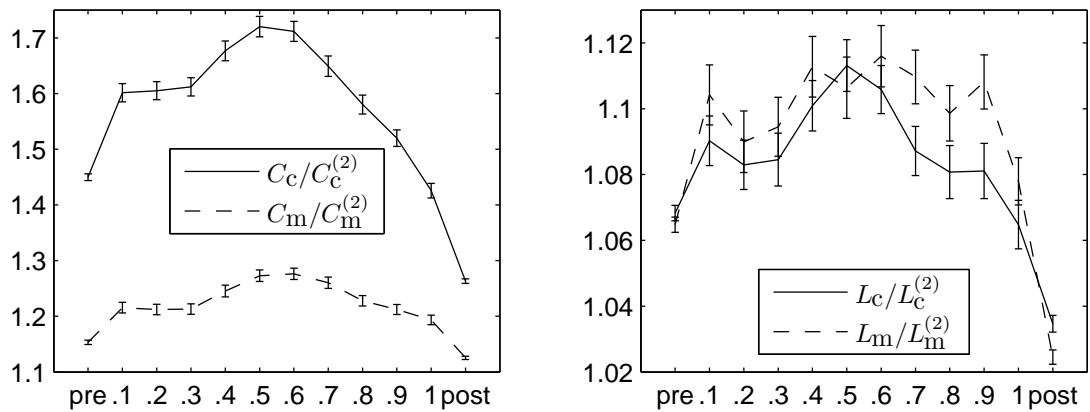}
\caption{{\bf Evolving normalized network properties averaged over 100 epileptic seizures.} Mean values of $C_{\mbox{c}}/C_{\mbox{c}}^{(2)}$ and $C_{\mbox{m}}/C_{\mbox{m}}^{(2)}$ (left) as well as $L_{\mbox{c}}/L_{\mbox{c}}^{(2)}$ and $L_{\mbox{m}}/L_{\mbox{m}}^{(2)}$ (right) averaged separately for pre-seizure, discretized seizure, and post-seizure time periods of 100 epileptic seizures. All error bars indicate standard error of the mean. Lines are for eye-guidance only.}
\label{fig:08}
\end{figure*}

In figure~\ref{fig:07} we show the mean values of properties of networks in each time bin for all seizures.
We observe \bCCa{}, \bCCb{}, \bLCa{}, \bLCb{}, \bCMa{}, and \bLMa{} to decrease during seizures and to increase prior to seizure end thereby roughly reflecting the amount of power contained in low frequencies (cf. figure~\ref{fig:09} (right), $\bar{P}_\delta$). \bCCa{} and \bCCb{} and to a lesser extent also \bLCa{} and \bLCb{} roughly follow the same course in time, however, with a slight shift in the range of values as already observed in the exemplary recording of a seizure (cf. figure~\ref{fig:05}). Differences between both random network ensembles are most pronounced in network properties based on \bfRm{}, i.e., between \bCMa{} and \bCMb{} as well as between \bLMa{} and \bLMb{}. This corroborates the observation that the clustering coefficient and the average shortest path length of the random networks based on \bfRm{} depend more sensitively on the power contents of EEG time series recorded from different brain regions than the respective quantities derived from \bfR{}. While \bLCO{} and \bLMO{} show a similar course in time, reaching a maximum in the middle of the seizures, we observe a remarkable difference between \bCCO{} and \bCMO{} prior to end of the seizures, where the amount of power in low frequencies is large. While \bCMO{} decreases at the end of the seizures, \bCCO{} does not and remains elevated after seizures. Interestingly, considering the corresponding quantities obtained from the second random network ensemble, \bCMb{} fluctuates around $0.3\pm 0.01$ and does not increase at the end of seizures, while, in contrast, \bCCb{} increases at the end of the seizures, traversing an interval of values roughly three times larger than the interval containing values of \bCMb{}. Taken together these findings suggest that the pronounced changes of the frequency content of EEG time series seen during epileptic seizures influence the values of the clustering coefficient and the average shortest path length.

A comparison of some value of a network property with the one obtained for a random network with the same edge density and number of nodes is typically achieved by calculating their ratio. If ER networks are used for comparison, the value of a network property is rescaled by a constant factor. In this case, the time-dependent changes of network properties shown in figure~\ref{fig:07} will be shifted along the ordinate only. In order to take into account the varying power content of EEG time series recorded from different brain regions we instead normalize the clustering coefficients and the average shortest path lengths with the corresponding quantities from the second random network ensemble \bCCb{}, \bCMb{}, \bLCb{}, and \bLMb{} (cf. figure~\ref{fig:08}). We observe the normalized network properties to describe a concave-like movement over time indicating a reconfiguration of networks from more random (before seizures) towards a more regular (during seizures) and back towards more random network topologies. This is in agreement with previous observations using a different and seldom used thresholding method \cite{Schindler2008a}. 

\section*{Discussion}

The network approach towards the analysis of empirical multivariate time series is based on the assumption that the data is well represented by a model of mutual relationships (i.e., a network). We studied interaction networks derived from finite time series generated by independent processes that would not advocate a representation by a model of mutual relationships. We observed the derived interaction networks to show non-trivial network topologies. These are induced by the finiteness of data, which limits reliability of estimators of signal interdependence, together with the use of a frequently employed thresholding technique. Since the analysis methodology alone can already introduce non-trivial structure in the derived networks, the question arises as to how informative network analysis results obtained from finite empirical data are with respect to the studied dynamics. This question may be addressed by defining and making use of appropriate null models. In the following, we briefly discuss two null models that are frequently employed in field studies.

\ER{} (ER) networks represent one of the earliest and best studied network models in mathematical literature and can be easily generated. They can be used to test whether the network under consideration complies with the notion of a random network in which possible edges are equally likely and independently chosen to become edges. We observed that clustering coefficient $C$ and average shortest path length $L$ for interaction networks derived from finite random time series differed pronouncedly from those obtained from corresponding ER networks, which would likely lead to a classification of interaction networks as small-world networks. Since the influence of the analysis methodology is not taken into account with ER networks, they may not be well suited for serving as null models in studies of interaction networks derived from finite time series.

Another null model is based on randomizing the network topology while preserving the degrees of nodes \cite{Roberts2000,Maslov2002,Maslov2004}. It is used to evaluate whether the network under consideration is random under the constraint of a given degree sequence. Results of our simulation studies point out that the structures induced in the network topology by the way how networks are derived from empirical time series cannot be related to the degree sequence only. We observed that $C$ and $L$ from interaction networks remarkably depended on the finiteness of the data, while the degree distribution did not (cf. figure~\ref{fig:03} (a-c), $M=1$). The usefulness of degree-preserving randomized networks has also been subject of debate since they do not take into account different characteristics of the data and its acquisition \cite{Randrup2004,Milo2004b}. Moreover, the link-switching algorithm frequently employed for generating such networks has been shown to non-uniformly sample the space of networks with predefined degree sequence (see, e.g., references \cite{Rao1996,Randrup2005}). This deficiency can be addressed by using alternative randomization schemes (see, e.g., \cite{Randrup2005,DelGenio2010,Blitzstein2010} and references therein).

We propose to take into account the finite length and the frequency contents of time series when defining null models. For this purpose we applied the same methodological steps as in field data analysis (estimation of signal interdependence and thresholding of interdependence values to define links) but used surrogate time series \cite{Schreiber2000a} to derive random networks (second ensemble). These surrogate time series comply with the null hypothesis of independent linear stochastic processes and preserve length, frequency content, and amplitude distribution of the original time series.  For these random networks, we observed (in our simulation studies) dependencies between properties of networks and properties of time series: the clustering coefficient $C$, and, to a lesser extent, the average shortest path length $L$ are higher the higher the relative amount of low frequency components, the shorter the length of time series, and the smaller the edge density of the network. Results obtained from an analysis of interaction networks derived from multichannel EEG recordings of one hundred epileptic seizures confirm that the pronounced changes of the frequency content seen during seizures influence the values of $C$ and $L$. Comparing these network characteristics with those obtained from our random networks allowed us to distinguish aspects of global network dynamics during seizures from those spuriously induced by the applied methods of analysis.

Our random networks will likely be classified as small-world networks when compared to ER networks which might indicate that small-world topologies in networks derived from empirical data as reported in an ever increasing number of studies can partly or solely be related to the finite length and frequency content of time series. If so, small-world topologies would be an overly complicated description of the simple finding of finite time series with a large amount of low frequency components. In this context, our approach could be of particular interest for studies that deal with short time series and low frequency contents, as, for example, is the case in resting state functional magnetic resonance imaging studies (see, e.g., references \cite{Eguiluz2005,VandenHeuvel2008,Hayasaka2010,Fransson2011,Tian2011}). In such studies, taking into account potential frequency effects could help to unravel information on the network level that would be otherwise masked.

We observed the degrees of nodes of our random networks to be correlated with the relative amount of power in low-frequencies in the respective time series (cf. figure~\ref{fig:03}). The degree of a node has been used in field studies as an indicator of its centrality in the network (see, e.g., \cite{Boccaletti2006a,Guye2010} and references therein). Particular interest has been devoted to nodes which are highly central (hubs). In this context it would be interesting to study whether findings of hubs in interaction networks
can partly or solely be explained by the various frequency contents of time series entering the analysis. In such a case, hubs would be a complicated representation of features already present on a single time series level. 
We are confident that our random networks can help to clarify this issue.

Our simulation studies were based on the simplified assumption that power spectra of all time series from which a network is derived are approximately equal. The dependencies of $C$ and $L$ on the power content could also be observed qualitatively for networks derived from EEG time series -- that were recorded from different brain regions and whose power spectra may differ substantially among each other -- but only if link definition was based on thresholding the values of the correlation coefficient ($\rho^\mathrm{c}$). Thus, estimating mean power spectra of multivariate time series can provide the experimentalist with a rule of thumb for the potential relative increase of $C$ and $L$ in different networks based on the correlation coefficient. This rule of thumb, however, might not be helpful if the maximum value of the absolute cross correlation ($\rho^\mathrm{m}$) is used to estimate signal interdependencies. In this case, $C$ and $L$ depended sensitively on the heterogeneity of power spectra (see the second random network ensemble). It would be interesting to investigate in future studies, which particular properties of $\rho^\mathrm{c}$ and $\rho^\mathrm{m}$ can be accounted for these differences.

We close the discussion with two remarks, the first being of interest for experimentalists. Our findings also shed light on a network construction technique that relies on significance testing in order to decide upon defining a link or not \cite{Kramer2009}. For this purpose, a null distribution of a chosen estimator of signal interdependence ($\rho^\mathrm{m}$) is generated for each pair of time series and a link is established if the null hypothesis of independent processes generating the time series can be rejected at a predefined significance level. It was suggested in Ref. \cite{Kramer2009} to use a limited subset of time series in order to minimize computational burden when generating null distributions. Our findings indicate that networks constructed this way will yield an artificially increased number of false positive or of false negative links which will depend on the frequency contents of time series being part or not part of the subset. Our second remark is related to network modeling. By choosing some threshold and generating time series that satisfy the relation between the size of the moving average and the length of time series, networks can be generated which differ in their degree distributions but approximately equal in their clustering coefficient and average shortest path length. This property could be of value for future modeling studies.

To summarize, we have demonstrated that interaction networks, derived from finite time series via thresholding an estimate of signal interdependence, can exhibit non-trivial properties that solely reflect the mostly unavoidable finiteness of empirical data, which limits the reliability of signal interdependence estimators. Addressing these influences, we proposed random network models that take into account the way interaction networks are derived from the data. With an exemplary time-resolved analysis of the clustering coefficient $C$ and the average shortest path length $L$ of interaction networks derived from multichannel electroencephalographic recordings of one hundred epileptic seizures, we demonstrated that our random networks allow one to gain deeper insights into the global network dynamics during seizures. Here we concentrated on $C$ and $L$ but we also expect other network characteristics to be influenced by the methodologies used to derive interaction networks from empirical data. Analytical investigations of properties of our random networks and the development of formal tests for deviations from these networks may be regarded as promising topics for further studies. Other research directions are related to the framework we proposed to generate random networks from time series. For example, parts of the framework may be exchanged in order to study network construction methodologies other than thresholding (e.g., based on minimum spanning trees \cite{Mantegna1999} or based on allowing weighted links) or other widely used linear and nonlinear methods for estimating signal interdependence \cite{Brillinger1981,Pikovsky_Book2001,Hlavackova2007}. Other surrogate concepts \cite{Small2001,Breakspear2003b,Nakamura2005,Keylock2006,Suzuki2007,Romano2009} may allow for defining different random networks tailored to various purposes. We believe that research into network inference from time series and into random network models that incorporate knowledge about the way networks are derived from empirical data can decisively advance applied network science. This line of research can contribute to gain a better understanding of complex dynamical systems studied in various scientific fields.

\appendix
\section{}
\subsection*{Lemma 1}

\noindent
For every $i,j\in\left\{1,\ldots,N\right\}$ with $i\neq j$, we have the following limit of the probability distribution of the empirical correlation:
\begin{equation}
P\left(\sqrt{\frac{T}{g(M)}}\mbox{corr}(x_{i,M,T},x_{j,M,T})\leq x\right)\rightarrow\Phi(x)
\end{equation}
with
\begin{equation*}
 g(M)=\frac{2}{3}M+\frac{1}{3}\frac{1}{M}
\end{equation*}
as $T\rightarrow\infty$, where $\Phi$ denotes the cumulative distribution function of a standard normal random variable.

\noindent
\textbf{Proof.} In order to simplify the presentation, we write $y_{i,M,T}(t)=x_{i,M,T}(t)-\frac{1}{2}$, so that $Ey_{i,M,T}(t)=0$. First note that $y_{i,M,T}(t)$ is a $M$-dependent sequence, i.e. for $|s-t|>M$, $y_{i,M,T}(s)$ and $y_{i,M,T}(t)$ are independent. So we have that the covariance\\ 
\begin{equation*}
\mathrm{Cov}\left(y_{i,M,T}(1)y_{j,M,T}(1),y_{i,M,T}(t)y_{j,M,T}(t)\right)=0
\end{equation*}
for $T>M$. Additionally, 
\begin{multline}
\mathrm{Cov}\left(y_{i,M,T}(1)y_{j,M,T}(1),y_{i,M,T}(t)y_{j,M,T}(t)\right) =\\ \mathrm{Cov}\left(y_{i,M,T}(1),y_{i,M,T}(t)\right) \mathrm{Cov}\left(y_{j,M,T}(1),y_{j,M,T}(t)\right) 
\end{multline}
and $\mathrm{Cov}\left(z_i(s),z_i(t)\right)=\mathrm{Var}\left(z_i(1)\right)$ if $s=t$ and otherwise $\mathrm{Cov}\left(z_i(s),z_i(t)\right)=0$. For $1\leq t\leq M$, we obtain by the definition of the moving average and the independence of the underlying process $z_j(t)$, $t\in\mathbb{N}$ that
\begin{eqnarray}
\mathrm{Cov}\left(y_{i,M,T}(1)y_{j,M,T}(1),y_{i,M,T}(t)y_{j,M,T}(t)\right)\\
=\frac{1}{M^4}\left(\sum_{s=1}^{M-(t-1)}\mathrm{Var}\left(z_j(s)\right)\right)^2\\
=\frac{1}{M^4}(M-(t-1))^2\mathrm{Var}^2\left(z_i(1)\right).
\end{eqnarray}
By the central limit theorem for $M$-dependent random variables, see reference \cite{Hoeffding1948},
\begin{multline}\label{eq:A3}
\frac{1}{\sqrt{\mathrm{Var}\left(\frac{1}{T}\sum_{t=1}^{T}y_{i,M,T}(t)y_{j,M,T}(t)\right)}}\\
\times\frac{1}{T}\sum_{t=1}^{T}y_{i,M,T}(t)y_{j,M,T}(t)
\end{multline}
converges in distribution to a standard normal random varibale as $T\rightarrow\infty$. Furthermore, we have the following convergence for the variance as $T\rightarrow\infty$:
\begin{widetext}
\begin{multline}\label{eq:A4}
T\mathrm{Var}\left(\frac{1}{T}\sum_{t=1}^{T}y_{i,M,T}(t)y_{j,M,T}(t)\right)\\
\rightarrow\mathrm{Var}(y_{i,M,T}(1)y_{j,M,T}(1))+2\sum_{t=2}^{M}\mathrm{Cov}\left(y_{i,M,T}(1)y_{j,M,T}(1),y_{i,M,T}(t)y_{j,M,T}(t)\right)\\
=\left(\frac{1}{M^2}+\frac{2}{M^4}\sum_{t=2}^{M}(M-(t-1))^2\right)\mathrm{Var}^2\left(z_i(1)\right)=\frac{g(M)}{M^2}\mathrm{Var}^2\left(z_i(1)\right).
\end{multline}
\end{widetext}
The last equality follows easily by $\sum_{i=1}^{n}i^2=\frac{n(n+1)(2n+1)}{6}$. With the same central limit theorem, $\frac{1}{\sqrt{T}}\sum_{t=1}^Ty_{i,M,T}(t)$ converges to a normal limit, so $\frac{1}{T^{\frac{3}{4}}}\sum_{t=1}^Ty_{i,M,T}(t)\rightarrow0$ in probability and consequently
\begin{multline}\label{eq:A5}
\sqrt{T}\left(\frac{1}{T}\sum_{t=1}^Ty_{i,M,T}(t)\right)\left(\frac{1}{T}\sum_{t=1}^Ty_{j,M,T}(t)\right)\\ =\left(\frac{1}{T^{\frac{3}{4}}}\sum_{t=1}^Ty_{i,M,T}(t)\right)\left(\frac{1}{T^{\frac{3}{4}}}\sum_{t=1}^Ty_{j,M,T}(t)\right)\rightarrow 0
\end{multline}
in probability as $T\rightarrow\infty$. By similar arguments, we have that $\frac{1}{T}\sum_{t=1}^Ty_{i,M,T}^2(t)\rightarrow\mathrm{ Var}(y_{i,M,T}(1))=\frac{1}{M}\mathrm{Var}\left(z_i(1)\right)$ and $\frac{1}{T}\sum_{t=1}^Ty_{i,M,T}(t)\rightarrow0$, so we get
\begin{multline}\label{eq:A6}
\frac{1}{T}\sum_{t=1}^T(y_{i,M,T}(t)-\bar{y}_{i,M,T})^2=\\
\frac{1}{T}\sum_{t=1}^Ty_{i,M,T}^2(t)-\left(\frac{1}{T}\sum_{t=1}^Ty_{i,M,T}(t)\right)^2\\
\rightarrow\mathrm{Var}(y_{i,M,T}(1))=\frac{1}{M}\mathrm{Var}\left(z_i(1)\right).
\end{multline}
By Slutsky's theorem \cite{Slutsky1925} and with \eqref{eq:A3}, \eqref{eq:A4}, \eqref{eq:A5}, and \eqref{eq:A6}, we finally obtain that
\begin{widetext}
\begin{equation}
\sqrt{\frac{T}{g(M)}}\mbox{corr}(x_{i,M,T},x_{j,M,T})=\frac{\sqrt{T}\frac{1}{T}\sum_{t=1}^{T}y_{i,M,T}(t)y_{j,M,T}(t)-\sqrt{T}\left(\frac{1}{T}\sum_{t=1}^Ty_{i,M,T}(t)\right)\left(\frac{1}{T}\sum_{t=1}^Ty_{j,M,T}(t)\right)}{\sqrt{g(M)\frac{1}{T}\sum_{t=1}^T(y_{i,M,T}(t)-\bar{y}_{i,M,T})^2\frac{1}{T}\sum_{t=1}^T(y_{j,M,T}(t)-\bar{y}_{j,M,T})^2}}
\end{equation}
\end{widetext}
converges in distribution to a standard normal random variable as $T\rightarrow\infty$. This completes the proof.

\subsection*{Lemma 2} 

\noindent
For $T\rightarrow\infty$, $R\rightarrow\infty$
\begin{equation}
\hat{\epsilon}\left(\frac{\theta}{\sqrt{T_\mathrm{eff}(M)}},M,T\right)\rightarrow2\Phi(-\theta)
\end{equation}
in probability with $T_\mathrm{eff}(M)=\frac{T}{g(M)}$.

\noindent
\textbf{Proof.}
With Lemma 1, we have that
\begin{widetext}
\begin{multline}
E\left[H_{ij,M,T}^{(r)}\left(\frac{\theta}{\sqrt{T_\mathrm{eff}(M)}}\right)\right]=P\left(\rho_{ij,M,T}>\frac{\theta}{\sqrt{T_\mathrm{eff}(M)}}\right)\nonumber\\
=P\left(\mbox{corr}(x_{i,M,T},x_{j,M,T})>\frac{\theta}{\sqrt{T_\mathrm{eff}(M)}}\right)+P\left(\mbox{corr}(x_{i,M,T},x_{j,M,T})<\frac{-\theta}{\sqrt{T_\mathrm{eff}(M)}}\right)\nonumber\\
= P\left(\sqrt{\frac{T}{g(M)}}\rho_{ij,M,T}>\theta\right)+P\left(\sqrt{\frac{T}{g(M)}}\rho_{ij,M,T}<-\theta\right)\rightarrow2\Phi(-\theta)
\end{multline}
\end{widetext}
as $T\rightarrow\infty$. Furthermore, $H_{ij,M,T}^{(r)}$ is bounded by $0$ and $1$, so $\mathrm{Var}\left(H_{ij,M,T}^{(r)}\right)\leq\frac{1}{4}$. By the independence of the $R$ random networks 
\begin{multline}
\mathrm{Var} \left(\hat{\epsilon}\left(\frac{\theta}{\sqrt{T_\mathrm{eff}(M)}},M,T\right)\right)\\
=\frac{1}{R^2}\sum_{r=1}^R \mathrm{Var} \left(H_{ij,M,T}^{(r)}\left(\frac{\theta}{\sqrt{T_\mathrm{eff}(M)}}\right)\right)\leq \frac{1}{4R}\rightarrow 0
\end{multline}
as $R\rightarrow\infty$. The lemma follows with the Chebyshev inequality.

\begin{acknowledgments}
We thank Marie-Therese Kuhnert, Gerrit Ansmann, and Alexander Rothkegel for their helpful comments and Paula Daniliuc for proofreading the manuscript. MW and SB were supported by the German National Academic Foundation. SB and KL acknowledge support from the German Science Foundation (LE 660/4-2).
\end{acknowledgments}


\begin{thebibliography}{74}
\expandafter\ifx\csname natexlab\endcsname\relax\def\natexlab#1{#1}\fi
\expandafter\ifx\csname bibnamefont\endcsname\relax
  \def\bibnamefont#1{#1}\fi
\expandafter\ifx\csname bibfnamefont\endcsname\relax
  \def\bibfnamefont#1{#1}\fi
\expandafter\ifx\csname citenamefont\endcsname\relax
  \def\citenamefont#1{#1}\fi
\expandafter\ifx\csname url\endcsname\relax
  \def\url#1{\texttt{#1}}\fi
\expandafter\ifx\csname urlprefix\endcsname\relax\def\urlprefix{URL }\fi
\providecommand{\bibinfo}[2]{#2}
\providecommand{\eprint}[2][]{\url{#2}}

\bibitem[{\citenamefont{Newman}(2003)}]{Newman2003}
\bibinfo{author}{\bibfnamefont{M.~E.~J.} \bibnamefont{Newman}},
  \bibinfo{journal}{SIAM Rev.} \textbf{\bibinfo{volume}{45}},
  \bibinfo{pages}{167} (\bibinfo{year}{2003}).

\bibitem[{\citenamefont{Boccaletti et~al.}(2006)\citenamefont{Boccaletti,
  Latora, Moreno, Chavez, and Hwang}}]{Boccaletti2006a}
\bibinfo{author}{\bibfnamefont{S.}~\bibnamefont{Boccaletti}},
  \bibinfo{author}{\bibfnamefont{V.}~\bibnamefont{Latora}},
  \bibinfo{author}{\bibfnamefont{Y.}~\bibnamefont{Moreno}},
  \bibinfo{author}{\bibfnamefont{M.}~\bibnamefont{Chavez}}, \bibnamefont{and}
  \bibinfo{author}{\bibfnamefont{D.-U.} \bibnamefont{Hwang}},
  \bibinfo{journal}{Phys. Rep.} \textbf{\bibinfo{volume}{424}},
  \bibinfo{pages}{175} (\bibinfo{year}{2006}).

\bibitem[{\citenamefont{Arenas et~al.}(2008)\citenamefont{Arenas,
  D{\'i}az-Guilera, Kurths, Moreno, and Zhou}}]{Arenas2008}
\bibinfo{author}{\bibfnamefont{A.}~\bibnamefont{Arenas}},
  \bibinfo{author}{\bibfnamefont{A.}~\bibnamefont{D{\'i}az-Guilera}},
  \bibinfo{author}{\bibfnamefont{J.}~\bibnamefont{Kurths}},
  \bibinfo{author}{\bibfnamefont{Y.}~\bibnamefont{Moreno}}, \bibnamefont{and}
  \bibinfo{author}{\bibfnamefont{C.}~\bibnamefont{Zhou}},
  \bibinfo{journal}{Phys. Rep.} \textbf{\bibinfo{volume}{469}},
  \bibinfo{pages}{93} (\bibinfo{year}{2008}).

\bibitem[{\citenamefont{Barrat et~al.}(2008)\citenamefont{Barrat,
  Barth{\'e}lemy, and Vespignani}}]{BarratBook2008}
\bibinfo{author}{\bibfnamefont{A.}~\bibnamefont{Barrat}},
  \bibinfo{author}{\bibfnamefont{M.}~\bibnamefont{Barth{\'e}lemy}},
  \bibnamefont{and}
  \bibinfo{author}{\bibfnamefont{A.}~\bibnamefont{Vespignani}},
  \emph{\bibinfo{title}{Dynamical Processes on Complex Networks}}
  (\bibinfo{publisher}{Cambridge University Press}, \bibinfo{address}{New York,
  USA}, \bibinfo{year}{2008}).

\bibitem[{\citenamefont{Tsonis and Roebber}(2004)}]{Tsonis2004}
\bibinfo{author}{\bibfnamefont{A.~A.} \bibnamefont{Tsonis}} \bibnamefont{and}
  \bibinfo{author}{\bibfnamefont{P.~J.} \bibnamefont{Roebber}},
  \bibinfo{journal}{Physica~A} \textbf{\bibinfo{volume}{333}},
  \bibinfo{pages}{497} (\bibinfo{year}{2004}).

\bibitem[{\citenamefont{Yamasaki et~al.}(2008)\citenamefont{Yamasaki,
  Gozolchiani, and Havlin}}]{Yamasaki2008}
\bibinfo{author}{\bibfnamefont{K.}~\bibnamefont{Yamasaki}},
  \bibinfo{author}{\bibfnamefont{A.}~\bibnamefont{Gozolchiani}},
  \bibnamefont{and} \bibinfo{author}{\bibfnamefont{S.}~\bibnamefont{Havlin}},
  \bibinfo{journal}{Phys. Rev. Lett.} \textbf{\bibinfo{volume}{100}},
  \bibinfo{pages}{228501} (\bibinfo{year}{2008}).

\bibitem[{\citenamefont{Donges et~al.}(2009{\natexlab{a}})\citenamefont{Donges,
  Zou, Marwan, and Kurths}}]{Donges2009}
\bibinfo{author}{\bibfnamefont{J.~F.} \bibnamefont{Donges}},
  \bibinfo{author}{\bibfnamefont{Y.}~\bibnamefont{Zou}},
  \bibinfo{author}{\bibfnamefont{N.}~\bibnamefont{Marwan}}, \bibnamefont{and}
  \bibinfo{author}{\bibfnamefont{J.}~\bibnamefont{Kurths}},
  \bibinfo{journal}{Eur. Phys. J.--Spec. Top.} \textbf{\bibinfo{volume}{174}},
  \bibinfo{pages}{157} (\bibinfo{year}{2009}{\natexlab{a}}).

\bibitem[{\citenamefont{Tsonis et~al.}(2011)\citenamefont{Tsonis, Wang,
  Swanson, Rodrigues, and {da Fontura Costa}}}]{Tsonis2010}
\bibinfo{author}{\bibfnamefont{A.~A.} \bibnamefont{Tsonis}},
  \bibinfo{author}{\bibfnamefont{G.}~\bibnamefont{Wang}},
  \bibinfo{author}{\bibfnamefont{K.~L.} \bibnamefont{Swanson}},
  \bibinfo{author}{\bibfnamefont{F.~A.} \bibnamefont{Rodrigues}},
  \bibnamefont{and} \bibinfo{author}{\bibfnamefont{L.}~\bibnamefont{{da Fontura
  Costa}}}, \bibinfo{journal}{Clim. Dynam.} \textbf{\bibinfo{volume}{37}},
  \bibinfo{pages}{933} (\bibinfo{year}{2011}).

\bibitem[{\citenamefont{Steinhaeuser et~al.}(2011)\citenamefont{Steinhaeuser,
  Chawla, and Ganguly}}]{Steinhaeuser2011}
\bibinfo{author}{\bibfnamefont{K.}~\bibnamefont{Steinhaeuser}},
  \bibinfo{author}{\bibfnamefont{N.~V.} \bibnamefont{Chawla}},
  \bibnamefont{and} \bibinfo{author}{\bibfnamefont{A.~R.}
  \bibnamefont{Ganguly}}, \bibinfo{journal}{Statistical Analysis and Data
  Mining} \textbf{\bibinfo{volume}{4}}, \bibinfo{pages}{497}
  (\bibinfo{year}{2011}).

\bibitem[{\citenamefont{Abe and Suzuki}(2004)}]{Abe2004}
\bibinfo{author}{\bibfnamefont{S.}~\bibnamefont{Abe}} \bibnamefont{and}
  \bibinfo{author}{\bibfnamefont{N.}~\bibnamefont{Suzuki}},
  \bibinfo{journal}{Physica A} \textbf{\bibinfo{volume}{337}},
  \bibinfo{pages}{357} (\bibinfo{year}{2004}).

\bibitem[{\citenamefont{Abe and Suzuki}(2006)}]{Abe2006}
\bibinfo{author}{\bibfnamefont{S.}~\bibnamefont{Abe}} \bibnamefont{and}
  \bibinfo{author}{\bibfnamefont{N.}~\bibnamefont{Suzuki}},
  \bibinfo{journal}{Nonlinear Proc. Geoph.} \textbf{\bibinfo{volume}{13}},
  \bibinfo{pages}{145} (\bibinfo{year}{2006}).

\bibitem[{\citenamefont{Jim{\'e}nez et~al.}(2008)\citenamefont{Jim{\'e}nez,
  Tiampo, and Posadas}}]{Jimenez2008}
\bibinfo{author}{\bibfnamefont{A.}~\bibnamefont{Jim{\'e}nez}},
  \bibinfo{author}{\bibfnamefont{K.~F.} \bibnamefont{Tiampo}},
  \bibnamefont{and} \bibinfo{author}{\bibfnamefont{A.~M.}
  \bibnamefont{Posadas}}, \bibinfo{journal}{Nonlinear Proc. Geoph.}
  \textbf{\bibinfo{volume}{15}}, \bibinfo{pages}{389} (\bibinfo{year}{2008}).

\bibitem[{\citenamefont{{Krishna Mohan} and Revathi}(2011)}]{Mohan2011}
\bibinfo{author}{\bibfnamefont{T.~R.} \bibnamefont{{Krishna Mohan}}}
  \bibnamefont{and} \bibinfo{author}{\bibfnamefont{P.~G.}
  \bibnamefont{Revathi}}, \bibinfo{journal}{J. Seismol.}
  \textbf{\bibinfo{volume}{15}}, \bibinfo{pages}{71} (\bibinfo{year}{2011}).

\bibitem[{\citenamefont{Mantegna}(1999)}]{Mantegna1999}
\bibinfo{author}{\bibfnamefont{R.~N.} \bibnamefont{Mantegna}},
  \bibinfo{journal}{Eur. Phys. J. B} \textbf{\bibinfo{volume}{11}},
  \bibinfo{pages}{193} (\bibinfo{year}{1999}).

\bibitem[{\citenamefont{Onnela et~al.}(2004)\citenamefont{Onnela, Kaski, and
  Kertesz}}]{Onnela2004}
\bibinfo{author}{\bibfnamefont{J.~P.} \bibnamefont{Onnela}},
  \bibinfo{author}{\bibfnamefont{K.}~\bibnamefont{Kaski}}, \bibnamefont{and}
  \bibinfo{author}{\bibfnamefont{J.}~\bibnamefont{Kertesz}},
  \bibinfo{journal}{Eur. Phys. J. B} \textbf{\bibinfo{volume}{38}},
  \bibinfo{pages}{353} (\bibinfo{year}{2004}).

\bibitem[{\citenamefont{Boginski et~al.}(2005)\citenamefont{Boginski, Butenko,
  and Pardalos}}]{Boginski2005}
\bibinfo{author}{\bibfnamefont{V.}~\bibnamefont{Boginski}},
  \bibinfo{author}{\bibfnamefont{S.}~\bibnamefont{Butenko}}, \bibnamefont{and}
  \bibinfo{author}{\bibfnamefont{P.~M.} \bibnamefont{Pardalos}},
  \bibinfo{journal}{Comput. Stat. An.} \textbf{\bibinfo{volume}{48}},
  \bibinfo{pages}{431} (\bibinfo{year}{2005}).

\bibitem[{\citenamefont{Qiu et~al.}(2010)\citenamefont{Qiu, Zheng, and
  Chen}}]{Qiu2010}
\bibinfo{author}{\bibfnamefont{T.}~\bibnamefont{Qiu}},
  \bibinfo{author}{\bibfnamefont{B.}~\bibnamefont{Zheng}}, \bibnamefont{and}
  \bibinfo{author}{\bibfnamefont{G.}~\bibnamefont{Chen}}, \bibinfo{journal}{New
  J. Physics} \textbf{\bibinfo{volume}{12}}, \bibinfo{pages}{043057}
  (\bibinfo{year}{2010}).

\bibitem[{\citenamefont{{Emmert-Streib} and
  Dehmer}(2010{\natexlab{a}})}]{Emmert-Streib2010}
\bibinfo{author}{\bibfnamefont{F.}~\bibnamefont{{Emmert-Streib}}}
  \bibnamefont{and} \bibinfo{author}{\bibfnamefont{M.}~\bibnamefont{Dehmer}},
  \bibinfo{journal}{PLoS ONE} \textbf{\bibinfo{volume}{5}},
  \bibinfo{pages}{e12884} (\bibinfo{year}{2010}{\natexlab{a}}).

\bibitem[{\citenamefont{Re{\ij}neveld et~al.}(2007)\citenamefont{Re{\ij}neveld,
  Ponten, Berendse, and Stam}}]{Reijneveld2007}
\bibinfo{author}{\bibfnamefont{J.~C.} \bibnamefont{Re{\ij}neveld}},
  \bibinfo{author}{\bibfnamefont{S.~C.} \bibnamefont{Ponten}},
  \bibinfo{author}{\bibfnamefont{H.~W.} \bibnamefont{Berendse}},
  \bibnamefont{and} \bibinfo{author}{\bibfnamefont{C.~J.} \bibnamefont{Stam}},
  \bibinfo{journal}{Clin. Neurophysiol.} \textbf{\bibinfo{volume}{118}},
  \bibinfo{pages}{2317} (\bibinfo{year}{2007}).

\bibitem[{\citenamefont{Bullmore and Sporns}(2009)}]{Bullmore2009}
\bibinfo{author}{\bibfnamefont{E.}~\bibnamefont{Bullmore}} \bibnamefont{and}
  \bibinfo{author}{\bibfnamefont{O.}~\bibnamefont{Sporns}},
  \bibinfo{journal}{Nat. Rev. Neurosci.} \textbf{\bibinfo{volume}{10}},
  \bibinfo{pages}{186} (\bibinfo{year}{2009}).

\bibitem[{\citenamefont{Kramer et~al.}(2009)\citenamefont{Kramer, Eden, Cash,
  and Kolaczyk}}]{Kramer2009}
\bibinfo{author}{\bibfnamefont{M.~A.} \bibnamefont{Kramer}},
  \bibinfo{author}{\bibfnamefont{U.~T.} \bibnamefont{Eden}},
  \bibinfo{author}{\bibfnamefont{S.~S.} \bibnamefont{Cash}}, \bibnamefont{and}
  \bibinfo{author}{\bibfnamefont{E.~D.} \bibnamefont{Kolaczyk}},
  \bibinfo{journal}{Phys. Rev.~E} \textbf{\bibinfo{volume}{79}},
  \bibinfo{pages}{061916} (\bibinfo{year}{2009}).

\bibitem[{\citenamefont{Donges et~al.}(2009{\natexlab{b}})\citenamefont{Donges,
  Zou, Marwan, and Kurths}}]{Donges2009b}
\bibinfo{author}{\bibfnamefont{J.~F.} \bibnamefont{Donges}},
  \bibinfo{author}{\bibfnamefont{Y.}~\bibnamefont{Zou}},
  \bibinfo{author}{\bibfnamefont{N.}~\bibnamefont{Marwan}}, \bibnamefont{and}
  \bibinfo{author}{\bibfnamefont{J.}~\bibnamefont{Kurths}},
  \bibinfo{journal}{Europhys. Lett.} \textbf{\bibinfo{volume}{87}},
  \bibinfo{pages}{48007} (\bibinfo{year}{2009}{\natexlab{b}}).

\bibitem[{\citenamefont{{Emmert-Streib} and
  Dehmer}(2010{\natexlab{b}})}]{Emmert-Streib2010b}
\bibinfo{author}{\bibfnamefont{F.}~\bibnamefont{{Emmert-Streib}}}
  \bibnamefont{and} \bibinfo{author}{\bibfnamefont{M.}~\bibnamefont{Dehmer}},
  \bibinfo{journal}{Complexity} \textbf{\bibinfo{volume}{16}},
  \bibinfo{pages}{24} (\bibinfo{year}{2010}{\natexlab{b}}).

\bibitem[{\citenamefont{Erd\H{o}s and R\'{e}nyi}(1959)}]{Erdos1959}
\bibinfo{author}{\bibfnamefont{P.}~\bibnamefont{Erd\H{o}s}} \bibnamefont{and}
  \bibinfo{author}{\bibfnamefont{A.}~\bibnamefont{R\'{e}nyi}},
  \bibinfo{journal}{Publ. Math. Debrecen} \textbf{\bibinfo{volume}{6}},
  \bibinfo{pages}{290} (\bibinfo{year}{1959}).

\bibitem[{\citenamefont{Rao et~al.}(1996)\citenamefont{Rao, Jana, and
  Bandyopadhyay}}]{Rao1996}
\bibinfo{author}{\bibfnamefont{A.~R.} \bibnamefont{Rao}},
  \bibinfo{author}{\bibfnamefont{R.}~\bibnamefont{Jana}}, \bibnamefont{and}
  \bibinfo{author}{\bibfnamefont{S.}~\bibnamefont{Bandyopadhyay}},
  \bibinfo{journal}{Sankhya Ser. A} \textbf{\bibinfo{volume}{58}},
  \bibinfo{pages}{225} (\bibinfo{year}{1996}).

\bibitem[{\citenamefont{Maslov and Sneppen}(2002)}]{Maslov2002}
\bibinfo{author}{\bibfnamefont{S.}~\bibnamefont{Maslov}} \bibnamefont{and}
  \bibinfo{author}{\bibfnamefont{K.}~\bibnamefont{Sneppen}},
  \bibinfo{journal}{Science} \textbf{\bibinfo{volume}{296}},
  \bibinfo{pages}{910} (\bibinfo{year}{2002}).

\bibitem[{\citenamefont{James et~al.}(2009)\citenamefont{James, Croft, and
  Krause}}]{James2009}
\bibinfo{author}{\bibfnamefont{R.}~\bibnamefont{James}},
  \bibinfo{author}{\bibfnamefont{D.~P.} \bibnamefont{Croft}}, \bibnamefont{and}
  \bibinfo{author}{\bibfnamefont{J.}~\bibnamefont{Krause}},
  \bibinfo{journal}{Behav. Ecol. Sociobiol.} \textbf{\bibinfo{volume}{63}},
  \bibinfo{pages}{989} (\bibinfo{year}{2009}).

\bibitem[{\citenamefont{{Lima-Mendez} and {van
  Helden}}(2009)}]{Lima-Mendez2009}
\bibinfo{author}{\bibfnamefont{G.}~\bibnamefont{{Lima-Mendez}}}
  \bibnamefont{and} \bibinfo{author}{\bibfnamefont{J.}~\bibnamefont{{van
  Helden}}}, \bibinfo{journal}{Mol. Biosyst.} \textbf{\bibinfo{volume}{5}},
  \bibinfo{pages}{1482} (\bibinfo{year}{2009}).

\bibitem[{\citenamefont{Ioannides}(2007)}]{Ioannides2007}
\bibinfo{author}{\bibfnamefont{A.~A.} \bibnamefont{Ioannides}},
  \bibinfo{journal}{Curr. Opin. Neurobiol.} \textbf{\bibinfo{volume}{17}},
  \bibinfo{pages}{161} (\bibinfo{year}{2007}).

\bibitem[{\citenamefont{Butts}(2009)}]{Butts2009}
\bibinfo{author}{\bibfnamefont{C.~T.} \bibnamefont{Butts}},
  \bibinfo{journal}{Science} \textbf{\bibinfo{volume}{325}},
  \bibinfo{pages}{414} (\bibinfo{year}{2009}).

\bibitem[{\citenamefont{Bialonski et~al.}(2010)\citenamefont{Bialonski,
  Horstmann, and Lehnertz}}]{Bialonski2010}
\bibinfo{author}{\bibfnamefont{S.}~\bibnamefont{Bialonski}},
  \bibinfo{author}{\bibfnamefont{M.-T.} \bibnamefont{Horstmann}},
  \bibnamefont{and} \bibinfo{author}{\bibfnamefont{K.}~\bibnamefont{Lehnertz}},
  \bibinfo{journal}{Chaos} \textbf{\bibinfo{volume}{20}},
  \bibinfo{pages}{013134} (\bibinfo{year}{2010}).

\bibitem[{\citenamefont{Antiqueira et~al.}(2010)\citenamefont{Antiqueira,
  Rodrigues, {van Wijk}, {da F. Costa}, and Daffertshofer}}]{Antiqueira2010}
\bibinfo{author}{\bibfnamefont{L.}~\bibnamefont{Antiqueira}},
  \bibinfo{author}{\bibfnamefont{F.~A.} \bibnamefont{Rodrigues}},
  \bibinfo{author}{\bibfnamefont{B.~C.~M.} \bibnamefont{{van Wijk}}},
  \bibinfo{author}{\bibfnamefont{L.}~\bibnamefont{{da F. Costa}}},
  \bibnamefont{and}
  \bibinfo{author}{\bibfnamefont{A.}~\bibnamefont{Daffertshofer}},
  \bibinfo{journal}{NeuroImage} \textbf{\bibinfo{volume}{53}},
  \bibinfo{pages}{439} (\bibinfo{year}{2010}).

\bibitem[{\citenamefont{Gerhard et~al.}(2011)\citenamefont{Gerhard, Pipa, Lima,
  Neuenschwander, and Gerstner}}]{Gerhard2011}
\bibinfo{author}{\bibfnamefont{F.}~\bibnamefont{Gerhard}},
  \bibinfo{author}{\bibfnamefont{G.}~\bibnamefont{Pipa}},
  \bibinfo{author}{\bibfnamefont{B.}~\bibnamefont{Lima}},
  \bibinfo{author}{\bibfnamefont{S.}~\bibnamefont{Neuenschwander}},
  \bibnamefont{and} \bibinfo{author}{\bibfnamefont{W.}~\bibnamefont{Gerstner}},
  \bibinfo{journal}{Front. Comp. Neuroscience} \textbf{\bibinfo{volume}{5}},
  \bibinfo{pages}{4} (\bibinfo{year}{2011}).

\bibitem[{\citenamefont{Brillinger}(1981)}]{Brillinger1981}
\bibinfo{author}{\bibfnamefont{D.}~\bibnamefont{Brillinger}},
  \emph{\bibinfo{title}{Time Series: Data Analysis and Theory}}
  (\bibinfo{publisher}{Holden-Day}, \bibinfo{address}{San Francisco, USA},
  \bibinfo{year}{1981}).

\bibitem[{\citenamefont{Pikovsky et~al.}(2001)\citenamefont{Pikovsky,
  Rosenblum, and Kurths}}]{Pikovsky_Book2001}
\bibinfo{author}{\bibfnamefont{A.~S.} \bibnamefont{Pikovsky}},
  \bibinfo{author}{\bibfnamefont{M.~G.} \bibnamefont{Rosenblum}},
  \bibnamefont{and} \bibinfo{author}{\bibfnamefont{J.}~\bibnamefont{Kurths}},
  \emph{\bibinfo{title}{Synchronization: {A} universal concept in nonlinear
  sciences}} (\bibinfo{publisher}{Cambridge University Press},
  \bibinfo{address}{Cambridge, UK}, \bibinfo{year}{2001}).

\bibitem[{\citenamefont{Boccaletti et~al.}(2002)\citenamefont{Boccaletti,
  Kurths, Osipov, Valladares, and Zhou}}]{Boccaletti2002}
\bibinfo{author}{\bibfnamefont{S.}~\bibnamefont{Boccaletti}},
  \bibinfo{author}{\bibfnamefont{J.}~\bibnamefont{Kurths}},
  \bibinfo{author}{\bibfnamefont{G.}~\bibnamefont{Osipov}},
  \bibinfo{author}{\bibfnamefont{D.~L.} \bibnamefont{Valladares}},
  \bibnamefont{and} \bibinfo{author}{\bibfnamefont{C.~S.} \bibnamefont{Zhou}},
  \bibinfo{journal}{Phys. Rep.} \textbf{\bibinfo{volume}{366}},
  \bibinfo{pages}{1} (\bibinfo{year}{2002}).

\bibitem[{\citenamefont{Kantz and Schreiber}(2003)}]{Kantz2003}
\bibinfo{author}{\bibfnamefont{H.}~\bibnamefont{Kantz}} \bibnamefont{and}
  \bibinfo{author}{\bibfnamefont{T.}~\bibnamefont{Schreiber}},
  \emph{\bibinfo{title}{Nonlinear Time Series Analysis}}
  (\bibinfo{publisher}{Cambridge University Press},
  \bibinfo{address}{Cambridge, UK}, \bibinfo{year}{2003}),
  \bibinfo{edition}{2nd} ed.

\bibitem[{\citenamefont{Pereda et~al.}(2005)\citenamefont{Pereda, {Quian
  Quiroga}, and Bhattacharya}}]{Pereda2005}
\bibinfo{author}{\bibfnamefont{E.}~\bibnamefont{Pereda}},
  \bibinfo{author}{\bibfnamefont{R.}~\bibnamefont{{Quian Quiroga}}},
  \bibnamefont{and}
  \bibinfo{author}{\bibfnamefont{J.}~\bibnamefont{Bhattacharya}},
  \bibinfo{journal}{Prog. Neurobiol.} \textbf{\bibinfo{volume}{77}},
  \bibinfo{pages}{1} (\bibinfo{year}{2005}).

\bibitem[{\citenamefont{Hlav{\'a}{\v c}kov{\'a}-Schindler
  et~al.}(2007)\citenamefont{Hlav{\'a}{\v c}kov{\'a}-Schindler, Palu{\v s},
  Vejmelka, and Bhattacharya}}]{Hlavackova2007}
\bibinfo{author}{\bibfnamefont{K.}~\bibnamefont{Hlav{\'a}{\v
  c}kov{\'a}-Schindler}},
  \bibinfo{author}{\bibfnamefont{M.}~\bibnamefont{Palu{\v s}}},
  \bibinfo{author}{\bibfnamefont{M.}~\bibnamefont{Vejmelka}}, \bibnamefont{and}
  \bibinfo{author}{\bibfnamefont{J.}~\bibnamefont{Bhattacharya}},
  \bibinfo{journal}{Phys. Rep.} \textbf{\bibinfo{volume}{441}},
  \bibinfo{pages}{1} (\bibinfo{year}{2007}).

\bibitem[{\citenamefont{Lehnertz et~al.}(2009)\citenamefont{Lehnertz,
  Bialonski, Horstmann, Krug, Rothkegel, Staniek, and Wagner}}]{Lehnertz2009b}
\bibinfo{author}{\bibfnamefont{K.}~\bibnamefont{Lehnertz}},
  \bibinfo{author}{\bibfnamefont{S.}~\bibnamefont{Bialonski}},
  \bibinfo{author}{\bibfnamefont{M.-T.} \bibnamefont{Horstmann}},
  \bibinfo{author}{\bibfnamefont{D.}~\bibnamefont{Krug}},
  \bibinfo{author}{\bibfnamefont{A.}~\bibnamefont{Rothkegel}},
  \bibinfo{author}{\bibfnamefont{M.}~\bibnamefont{Staniek}}, \bibnamefont{and}
  \bibinfo{author}{\bibfnamefont{T.}~\bibnamefont{Wagner}},
  \bibinfo{journal}{J. Neurosci. Methods} \textbf{\bibinfo{volume}{183}},
  \bibinfo{pages}{42} (\bibinfo{year}{2009}).

\bibitem[{\citenamefont{Watts and Strogatz}(1998)}]{Watts1998}
\bibinfo{author}{\bibfnamefont{D.~J.} \bibnamefont{Watts}} \bibnamefont{and}
  \bibinfo{author}{\bibfnamefont{S.~H.} \bibnamefont{Strogatz}},
  \bibinfo{journal}{Nature} \textbf{\bibinfo{volume}{393}},
  \bibinfo{pages}{440} (\bibinfo{year}{1998}).

\bibitem[{\citenamefont{Latora and Marchiori}(2001)}]{Latora2001}
\bibinfo{author}{\bibfnamefont{V.}~\bibnamefont{Latora}} \bibnamefont{and}
  \bibinfo{author}{\bibfnamefont{M.}~\bibnamefont{Marchiori}},
  \bibinfo{journal}{Phys. Rev. Lett.} \textbf{\bibinfo{volume}{87}},
  \bibinfo{pages}{198701} (\bibinfo{year}{2001}).

\bibitem[{\citenamefont{Latora and Marchiori}(2003)}]{Latora2003}
\bibinfo{author}{\bibfnamefont{V.}~\bibnamefont{Latora}} \bibnamefont{and}
  \bibinfo{author}{\bibfnamefont{M.}~\bibnamefont{Marchiori}},
  \bibinfo{journal}{Eur. Phys. J. B} \textbf{\bibinfo{volume}{32}},
  \bibinfo{pages}{249} (\bibinfo{year}{2003}).

\bibitem[{\citenamefont{Chung and Lu}(2001)}]{Chung2001}
\bibinfo{author}{\bibfnamefont{F.}~\bibnamefont{Chung}} \bibnamefont{and}
  \bibinfo{author}{\bibfnamefont{L.}~\bibnamefont{Lu}}, \bibinfo{journal}{Adv.
  Appl. Math.} \textbf{\bibinfo{volume}{26}}, \bibinfo{pages}{257}
  (\bibinfo{year}{2001}).

\bibitem[{\citenamefont{Press et~al.}(2002)\citenamefont{Press, Teukolsky,
  Vetterling, and Flannery}}]{Press2002}
\bibinfo{author}{\bibfnamefont{W.~H.} \bibnamefont{Press}},
  \bibinfo{author}{\bibfnamefont{S.~A.} \bibnamefont{Teukolsky}},
  \bibinfo{author}{\bibfnamefont{W.~T.} \bibnamefont{Vetterling}},
  \bibnamefont{and} \bibinfo{author}{\bibfnamefont{B.~P.}
  \bibnamefont{Flannery}}, \emph{\bibinfo{title}{Numerical {R}ecipes in {C}}}
  (\bibinfo{publisher}{Cambridge University Press},
  \bibinfo{address}{Cambridge, UK}, \bibinfo{year}{2002}),
  \bibinfo{edition}{2nd} ed.

\bibitem[{\citenamefont{Franaszczuk et~al.}(1998)\citenamefont{Franaszczuk,
  Bergey, Durka, and Eisenberg}}]{Franaszczuk1998b}
\bibinfo{author}{\bibfnamefont{P.~J.} \bibnamefont{Franaszczuk}},
  \bibinfo{author}{\bibfnamefont{G.~K.} \bibnamefont{Bergey}},
  \bibinfo{author}{\bibfnamefont{P.~J.} \bibnamefont{Durka}}, \bibnamefont{and}
  \bibinfo{author}{\bibfnamefont{H.~M.} \bibnamefont{Eisenberg}},
  \bibinfo{journal}{Electroencephalogr. Clin. Neurophysiol.}
  \textbf{\bibinfo{volume}{106}}, \bibinfo{pages}{513} (\bibinfo{year}{1998}).

\bibitem[{\citenamefont{Schiff et~al.}(2000)\citenamefont{Schiff, Colella,
  Jacyna, Hughes, Creekmore, Marshall, {Bozek-Kuzmicki}, Benke, Gaillard, Conry
  et~al.}}]{Schiff2000}
\bibinfo{author}{\bibfnamefont{S.~J.} \bibnamefont{Schiff}},
  \bibinfo{author}{\bibfnamefont{D.}~\bibnamefont{Colella}},
  \bibinfo{author}{\bibfnamefont{G.~M.} \bibnamefont{Jacyna}},
  \bibinfo{author}{\bibfnamefont{E.}~\bibnamefont{Hughes}},
  \bibinfo{author}{\bibfnamefont{J.~W.} \bibnamefont{Creekmore}},
  \bibinfo{author}{\bibfnamefont{A.}~\bibnamefont{Marshall}},
  \bibinfo{author}{\bibfnamefont{M.}~\bibnamefont{{Bozek-Kuzmicki}}},
  \bibinfo{author}{\bibfnamefont{G.}~\bibnamefont{Benke}},
  \bibinfo{author}{\bibfnamefont{W.~D.} \bibnamefont{Gaillard}},
  \bibinfo{author}{\bibfnamefont{J.}~\bibnamefont{Conry}},
  \bibnamefont{et~al.}, \bibinfo{journal}{Clin. Neurophysiol.}
  \textbf{\bibinfo{volume}{111}}, \bibinfo{pages}{953} (\bibinfo{year}{2000}).

\bibitem[{\citenamefont{Jouny et~al.}(2003)\citenamefont{Jouny, Franaszczuk,
  and Bergey}}]{Jouny2003}
\bibinfo{author}{\bibfnamefont{C.~C.} \bibnamefont{Jouny}},
  \bibinfo{author}{\bibfnamefont{P.~J.} \bibnamefont{Franaszczuk}},
  \bibnamefont{and} \bibinfo{author}{\bibfnamefont{G.~K.}
  \bibnamefont{Bergey}}, \bibinfo{journal}{Clin. Neurophysiol.}
  \textbf{\bibinfo{volume}{114}}, \bibinfo{pages}{426} (\bibinfo{year}{2003}).

\bibitem[{\citenamefont{Bartolomei et~al.}(2010)\citenamefont{Bartolomei,
  {Cosandier-Rimele}, {McGonigal}, Aubert, Regis, Gavaret, Wendling, and
  Chauvel}}]{Bartolomei2010}
\bibinfo{author}{\bibfnamefont{F.}~\bibnamefont{Bartolomei}},
  \bibinfo{author}{\bibfnamefont{D.}~\bibnamefont{{Cosandier-Rimele}}},
  \bibinfo{author}{\bibfnamefont{A.}~\bibnamefont{{McGonigal}}},
  \bibinfo{author}{\bibfnamefont{S.}~\bibnamefont{Aubert}},
  \bibinfo{author}{\bibfnamefont{J.}~\bibnamefont{Regis}},
  \bibinfo{author}{\bibfnamefont{M.}~\bibnamefont{Gavaret}},
  \bibinfo{author}{\bibfnamefont{F.}~\bibnamefont{Wendling}}, \bibnamefont{and}
  \bibinfo{author}{\bibfnamefont{P.}~\bibnamefont{Chauvel}},
  \bibinfo{journal}{Epilepsia} \textbf{\bibinfo{volume}{51}},
  \bibinfo{pages}{2147} (\bibinfo{year}{2010}).

\bibitem[{\citenamefont{Schindler et~al.}(2008)\citenamefont{Schindler,
  Bialonski, Horstmann, Elger, and Lehnertz}}]{Schindler2008a}
\bibinfo{author}{\bibfnamefont{K.}~\bibnamefont{Schindler}},
  \bibinfo{author}{\bibfnamefont{S.}~\bibnamefont{Bialonski}},
  \bibinfo{author}{\bibfnamefont{M.-T.} \bibnamefont{Horstmann}},
  \bibinfo{author}{\bibfnamefont{C.~E.} \bibnamefont{Elger}}, \bibnamefont{and}
  \bibinfo{author}{\bibfnamefont{K.}~\bibnamefont{Lehnertz}},
  \bibinfo{journal}{Chaos} \textbf{\bibinfo{volume}{18}},
  \bibinfo{pages}{033119} (\bibinfo{year}{2008}).

\bibitem[{\citenamefont{Schindler et~al.}(2007)\citenamefont{Schindler, Leung,
  Elger, and Lehnertz}}]{Schindler2007a}
\bibinfo{author}{\bibfnamefont{K.}~\bibnamefont{Schindler}},
  \bibinfo{author}{\bibfnamefont{H.}~\bibnamefont{Leung}},
  \bibinfo{author}{\bibfnamefont{C.~E.} \bibnamefont{Elger}}, \bibnamefont{and}
  \bibinfo{author}{\bibfnamefont{K.}~\bibnamefont{Lehnertz}},
  \bibinfo{journal}{Brain} \textbf{\bibinfo{volume}{130}}, \bibinfo{pages}{65}
  (\bibinfo{year}{2007}).

\bibitem[{\citenamefont{Schreiber and Schmitz}(1996)}]{Schreiber1996a}
\bibinfo{author}{\bibfnamefont{T.}~\bibnamefont{Schreiber}} \bibnamefont{and}
  \bibinfo{author}{\bibfnamefont{A.}~\bibnamefont{Schmitz}},
  \bibinfo{journal}{Phys. Rev. Lett.} \textbf{\bibinfo{volume}{77}},
  \bibinfo{pages}{635} (\bibinfo{year}{1996}).

\bibitem[{\citenamefont{Schreiber and Schmitz}(2000)}]{Schreiber2000a}
\bibinfo{author}{\bibfnamefont{T.}~\bibnamefont{Schreiber}} \bibnamefont{and}
  \bibinfo{author}{\bibfnamefont{A.}~\bibnamefont{Schmitz}},
  \bibinfo{journal}{Physica~D} \textbf{\bibinfo{volume}{142}},
  \bibinfo{pages}{346} (\bibinfo{year}{2000}).

\bibitem[{\citenamefont{Roberts}(2000)}]{Roberts2000}
\bibinfo{author}{\bibfnamefont{J.~M.} \bibnamefont{Roberts}},
  \bibinfo{journal}{Soc. Networks} \textbf{\bibinfo{volume}{22}},
  \bibinfo{pages}{273} (\bibinfo{year}{2000}).

\bibitem[{\citenamefont{Maslov et~al.}(2004)\citenamefont{Maslov, Sneppen, and
  Zaliznyak}}]{Maslov2004}
\bibinfo{author}{\bibfnamefont{S.}~\bibnamefont{Maslov}},
  \bibinfo{author}{\bibfnamefont{K.}~\bibnamefont{Sneppen}}, \bibnamefont{and}
  \bibinfo{author}{\bibfnamefont{A.}~\bibnamefont{Zaliznyak}},
  \bibinfo{journal}{Physica~A} \textbf{\bibinfo{volume}{333}},
  \bibinfo{pages}{529} (\bibinfo{year}{2004}).

\bibitem[{\citenamefont{{Artzy-Randrup}
  et~al.}(2004)\citenamefont{{Artzy-Randrup}, Fleishman, {Ben-Tal}, and
  Stone}}]{Randrup2004}
\bibinfo{author}{\bibfnamefont{Y.}~\bibnamefont{{Artzy-Randrup}}},
  \bibinfo{author}{\bibfnamefont{S.~J.} \bibnamefont{Fleishman}},
  \bibinfo{author}{\bibfnamefont{N.}~\bibnamefont{{Ben-Tal}}},
  \bibnamefont{and} \bibinfo{author}{\bibfnamefont{L.}~\bibnamefont{Stone}},
  \bibinfo{journal}{Science} \textbf{\bibinfo{volume}{305}},
  \bibinfo{pages}{1107} (\bibinfo{year}{2004}).

\bibitem[{\citenamefont{Milo et~al.}(2004)\citenamefont{Milo, Itzkovitz,
  Kashtan, Levitt, and Alon}}]{Milo2004b}
\bibinfo{author}{\bibfnamefont{R.}~\bibnamefont{Milo}},
  \bibinfo{author}{\bibfnamefont{S.}~\bibnamefont{Itzkovitz}},
  \bibinfo{author}{\bibfnamefont{N.}~\bibnamefont{Kashtan}},
  \bibinfo{author}{\bibfnamefont{R.}~\bibnamefont{Levitt}}, \bibnamefont{and}
  \bibinfo{author}{\bibfnamefont{U.}~\bibnamefont{Alon}},
  \bibinfo{journal}{Science} \textbf{\bibinfo{volume}{305}},
  \bibinfo{pages}{1107} (\bibinfo{year}{2004}).

\bibitem[{\citenamefont{{Artzy-Randrup} and Stone}(2005)}]{Randrup2005}
\bibinfo{author}{\bibfnamefont{Y.}~\bibnamefont{{Artzy-Randrup}}}
  \bibnamefont{and} \bibinfo{author}{\bibfnamefont{L.}~\bibnamefont{Stone}},
  \bibinfo{journal}{Phys. Rev. E} \textbf{\bibinfo{volume}{72}},
  \bibinfo{pages}{056708} (\bibinfo{year}{2005}).

\bibitem[{\citenamefont{{Del Genio} et~al.}(2010)\citenamefont{{Del Genio},
  Kim, Toroczkai, and Bassler}}]{DelGenio2010}
\bibinfo{author}{\bibfnamefont{C.~I.} \bibnamefont{{Del Genio}}},
  \bibinfo{author}{\bibfnamefont{H.}~\bibnamefont{Kim}},
  \bibinfo{author}{\bibfnamefont{Z.}~\bibnamefont{Toroczkai}},
  \bibnamefont{and} \bibinfo{author}{\bibfnamefont{K.~E.}
  \bibnamefont{Bassler}}, \bibinfo{journal}{PLoS ONE}
  \textbf{\bibinfo{volume}{5}}, \bibinfo{pages}{e10012} (\bibinfo{year}{2010}).

\bibitem[{\citenamefont{Blitzstein and Diaconis}(2010)}]{Blitzstein2010}
\bibinfo{author}{\bibfnamefont{J.}~\bibnamefont{Blitzstein}} \bibnamefont{and}
  \bibinfo{author}{\bibfnamefont{P.}~\bibnamefont{Diaconis}},
  \bibinfo{journal}{Internet Mathematics} \textbf{\bibinfo{volume}{6}},
  \bibinfo{pages}{489} (\bibinfo{year}{2010}).

\bibitem[{\citenamefont{Eguiluz et~al.}(2005)\citenamefont{Eguiluz, Chialvo,
  Cecchi, Baliki, and Apkarian}}]{Eguiluz2005}
\bibinfo{author}{\bibfnamefont{V.~M.} \bibnamefont{Eguiluz}},
  \bibinfo{author}{\bibfnamefont{D.~R.} \bibnamefont{Chialvo}},
  \bibinfo{author}{\bibfnamefont{G.~A.} \bibnamefont{Cecchi}},
  \bibinfo{author}{\bibfnamefont{M.}~\bibnamefont{Baliki}}, \bibnamefont{and}
  \bibinfo{author}{\bibfnamefont{A.~V.} \bibnamefont{Apkarian}},
  \bibinfo{journal}{Phys. Rev. Lett.} \textbf{\bibinfo{volume}{94}},
  \bibinfo{pages}{018102} (\bibinfo{year}{2005}).

\bibitem[{\citenamefont{{van den Heuvel} et~al.}(2008)\citenamefont{{van den
  Heuvel}, Stam, Boersma, and {Hulshoff Pol}}}]{VandenHeuvel2008}
\bibinfo{author}{\bibfnamefont{M.~P.} \bibnamefont{{van den Heuvel}}},
  \bibinfo{author}{\bibfnamefont{C.~J.} \bibnamefont{Stam}},
  \bibinfo{author}{\bibfnamefont{M.}~\bibnamefont{Boersma}}, \bibnamefont{and}
  \bibinfo{author}{\bibfnamefont{H.~E.} \bibnamefont{{Hulshoff Pol}}},
  \bibinfo{journal}{NeuroImage} \textbf{\bibinfo{volume}{43}},
  \bibinfo{pages}{528} (\bibinfo{year}{2008}).

\bibitem[{\citenamefont{Hayasaka and Laurienti}(2010)}]{Hayasaka2010}
\bibinfo{author}{\bibfnamefont{S.}~\bibnamefont{Hayasaka}} \bibnamefont{and}
  \bibinfo{author}{\bibfnamefont{P.~J.} \bibnamefont{Laurienti}},
  \bibinfo{journal}{NeuroImage} \textbf{\bibinfo{volume}{50}},
  \bibinfo{pages}{499} (\bibinfo{year}{2010}).

\bibitem[{\citenamefont{Fransson et~al.}(2011)\citenamefont{Fransson, \r{A}den,
  Blennow, and Lagercrantz}}]{Fransson2011}
\bibinfo{author}{\bibfnamefont{P.}~\bibnamefont{Fransson}},
  \bibinfo{author}{\bibfnamefont{U.}~\bibnamefont{\r{A}den}},
  \bibinfo{author}{\bibfnamefont{M.}~\bibnamefont{Blennow}}, \bibnamefont{and}
  \bibinfo{author}{\bibfnamefont{H.}~\bibnamefont{Lagercrantz}},
  \bibinfo{journal}{Cereb. Cortex} \textbf{\bibinfo{volume}{21}},
  \bibinfo{pages}{145} (\bibinfo{year}{2011}).

\bibitem[{\citenamefont{Tian et~al.}(2011)\citenamefont{Tian, Wang, Yan, and
  He}}]{Tian2011}
\bibinfo{author}{\bibfnamefont{L.}~\bibnamefont{Tian}},
  \bibinfo{author}{\bibfnamefont{J.}~\bibnamefont{Wang}},
  \bibinfo{author}{\bibfnamefont{C.}~\bibnamefont{Yan}}, \bibnamefont{and}
  \bibinfo{author}{\bibfnamefont{Y.}~\bibnamefont{He}},
  \bibinfo{journal}{NeuroImage} \textbf{\bibinfo{volume}{54}},
  \bibinfo{pages}{191} (\bibinfo{year}{2011}).

\bibitem[{\citenamefont{Guye et~al.}(2010)\citenamefont{Guye, Bettus,
  Bartolomei, and Cozzone}}]{Guye2010}
\bibinfo{author}{\bibfnamefont{M.}~\bibnamefont{Guye}},
  \bibinfo{author}{\bibfnamefont{G.}~\bibnamefont{Bettus}},
  \bibinfo{author}{\bibfnamefont{F.}~\bibnamefont{Bartolomei}},
  \bibnamefont{and} \bibinfo{author}{\bibfnamefont{P.~J.}
  \bibnamefont{Cozzone}}, \bibinfo{journal}{Magn. Reson. Mater. Phy.}
  \textbf{\bibinfo{volume}{23}}, \bibinfo{pages}{409} (\bibinfo{year}{2010}).

\bibitem[{\citenamefont{Small et~al.}(2001)\citenamefont{Small, Yu, and
  Harrison}}]{Small2001}
\bibinfo{author}{\bibfnamefont{M.}~\bibnamefont{Small}},
  \bibinfo{author}{\bibfnamefont{D.}~\bibnamefont{Yu}}, \bibnamefont{and}
  \bibinfo{author}{\bibfnamefont{R.~G.} \bibnamefont{Harrison}},
  \bibinfo{journal}{Phys. Rev. Lett.} \textbf{\bibinfo{volume}{87}},
  \bibinfo{pages}{188101} (\bibinfo{year}{2001}).

\bibitem[{\citenamefont{Breakspear et~al.}(2003)\citenamefont{Breakspear,
  Brammer, and Robinson}}]{Breakspear2003b}
\bibinfo{author}{\bibfnamefont{M.}~\bibnamefont{Breakspear}},
  \bibinfo{author}{\bibfnamefont{M.}~\bibnamefont{Brammer}}, \bibnamefont{and}
  \bibinfo{author}{\bibfnamefont{P.~A.} \bibnamefont{Robinson}},
  \bibinfo{journal}{Physica~D} \textbf{\bibinfo{volume}{182}},
  \bibinfo{pages}{1} (\bibinfo{year}{2003}).

\bibitem[{\citenamefont{Nakamura and Small}(2005)}]{Nakamura2005}
\bibinfo{author}{\bibfnamefont{T.}~\bibnamefont{Nakamura}} \bibnamefont{and}
  \bibinfo{author}{\bibfnamefont{M.}~\bibnamefont{Small}},
  \bibinfo{journal}{Phys. Rev. E} \textbf{\bibinfo{volume}{72}},
  \bibinfo{pages}{056216} (\bibinfo{year}{2005}).

\bibitem[{\citenamefont{Keylock}(2006)}]{Keylock2006}
\bibinfo{author}{\bibfnamefont{C.~J.} \bibnamefont{Keylock}},
  \bibinfo{journal}{Phys. Rev. E} \textbf{\bibinfo{volume}{73}},
  \bibinfo{pages}{036707} (\bibinfo{year}{2006}).

\bibitem[{\citenamefont{Suzuki et~al.}(2007)\citenamefont{Suzuki, Ikeguchi, and
  Suzuki}}]{Suzuki2007}
\bibinfo{author}{\bibfnamefont{T.}~\bibnamefont{Suzuki}},
  \bibinfo{author}{\bibfnamefont{T.}~\bibnamefont{Ikeguchi}}, \bibnamefont{and}
  \bibinfo{author}{\bibfnamefont{M.}~\bibnamefont{Suzuki}},
  \bibinfo{journal}{Physica D} \textbf{\bibinfo{volume}{231}},
  \bibinfo{pages}{108} (\bibinfo{year}{2007}).

\bibitem[{\citenamefont{Romano et~al.}(2009)\citenamefont{Romano, Thiel,
  Kurths, Mergenthaler, and Engbert}}]{Romano2009}
\bibinfo{author}{\bibfnamefont{M.~C.} \bibnamefont{Romano}},
  \bibinfo{author}{\bibfnamefont{M.}~\bibnamefont{Thiel}},
  \bibinfo{author}{\bibfnamefont{J.}~\bibnamefont{Kurths}},
  \bibinfo{author}{\bibfnamefont{K.}~\bibnamefont{Mergenthaler}},
  \bibnamefont{and} \bibinfo{author}{\bibfnamefont{R.}~\bibnamefont{Engbert}},
  \bibinfo{journal}{Chaos} \textbf{\bibinfo{volume}{19}},
  \bibinfo{pages}{015108} (\bibinfo{year}{2009}).

\bibitem[{\citenamefont{Hoeffding and Robbins}(1948)}]{Hoeffding1948}
\bibinfo{author}{\bibfnamefont{W.}~\bibnamefont{Hoeffding}} \bibnamefont{and}
  \bibinfo{author}{\bibfnamefont{H.}~\bibnamefont{Robbins}},
  \bibinfo{journal}{Duke Math. J.} \textbf{\bibinfo{volume}{15}},
  \bibinfo{pages}{773} (\bibinfo{year}{1948}).

\bibitem[{\citenamefont{Slutsky}(1925)}]{Slutsky1925}
\bibinfo{author}{\bibfnamefont{E.}~\bibnamefont{Slutsky}},
  \bibinfo{journal}{Metron} \textbf{\bibinfo{volume}{5}}, \bibinfo{pages}{3}
  (\bibinfo{year}{1925}).

\end{thebibliography}
\end{document}